\documentclass[12pt]{article}

\usepackage{setspace}

\usepackage{amsmath}
\usepackage{graphicx}
\usepackage{natbib}
\usepackage{url} % not crucial - just used below for the URL 
\usepackage{latexsym}
\usepackage{amsfonts}
\usepackage{amssymb}
\usepackage{psfrag}
\usepackage{graphicx}
\usepackage{calc, cancel}
\usepackage{graphics, graphicx, caption, subcaption}
\usepackage{latexsym, pgfplots, tikz, url,verbatim,multicol,enumerate,bm,listings,color,hyperref}
\usepackage{multirow}
\usepackage[justification=centering]{caption}

\usepackage{wrapfig, sidecap,enumitem}
\usepackage{epsfig}
\usepackage{longtable}

\newcommand{\normal}{\mathop{\rm N}}

\newcommand{\GIG}{\mathop{\rm GIG}}

\newcommand{\Lc}{\mathcal{L}}

\newcommand{\sumn}{\sum_{i = 1}^{n}}

\newcommand{\betav}{\mbox{\boldmath{$\beta$}}}

\newcommand{\muv}{\mbox{\boldmath{$\mu$}}}

\newcommand{\Sigmav}{\mbox{\boldmath{$\Sigma$}}}
\newcommand{\Deltav}{\mbox{\boldmath{$\Delta$}}}

\newcommand{\thetav}{\mbox{\boldmath{$\theta$}}}

\def\half{\frac{1}{2}}

\def\Id{\mathbb I} % identity matrix

\def\Yv{\mathbf Y}

\def\uv{\mathbf u}

\def\av{\mathbf a}

\def\yv{\mathbf y}
\def\zv{\mathbf z}
\def\tv{\mathbf t}

\def\1v{\mathbf 1}
\def\0v{\mathbf 0}

\begin{document}
\doublespacing

\title{A Bayesian approach for clustering skewed data using mixtures of multivariate normal-inverse Gaussian distributions}

\author{Yuan Fang\footnote{Department of Mathematical Sciences, Binghamton University, State University of New York, 4400 Vestal Parkway East, Binghamton, NY, USA 13902. e: yfang8@binghamton.edu} \and Dimitris Karlis \footnote{Department of Statistics, University of Waterloo, thens University of Economics and Business, Athens, Greece. e: karlis@aueb.gr} \and Sanjeena Subedi \footnote{Department of Mathematical Sciences, Binghamton University, State University of New York, 4400 Vestal Parkway East, Binghamton, NY, USA 13902. e: sdang@binghamton.edu}}

\maketitle

\begin{abstract}
Non-Gaussian mixture models are gaining increasing attention for
mixture model-based clustering particularly  when dealing with
data that exhibit features such as skewness and heavy tails. Here,
such a mixture distribution is presented, based on the
multivariate normal inverse Gaussian (MNIG) distribution. For parameter
estimation of the mixture, a Bayesian approach via Gibbs sampler
is used; for this, a novel approach to simulate univariate
generalized inverse Gaussian random variables and matrix
generalized inverse Gaussian random matrices is provided. The
proposed algorithm will be applied to both simulated and real
data. Through simulation studies and real data analysis, we show
parameter recovery and that our approach provides competitive
clustering results compared to other clustering approaches.
\end{abstract}

\textbf{Keywords}: GIG-distribution, cluster analysis, matrix-GIG distribution, model-based clustering, multivariate skew distributions, MNIG distribution.

\section{Introduction}
\label{sec:intro}

    Model-based clustering uses finite mixture models that assumes that the population consists
    of a finite mixture of subpopulations, each represented by a known distribution.
    Gaussian mixture models, where each of the mixture components comes from a Gaussian distribution
    are predominant in the literature. However, Gaussian mixture models can only model symmetric elliptical data.
    In the last decade, skewed mixture models that are based off of non-symmetric marginal distributions
     and have the flexibility of representing skewed and symmetric components have increasingly gained attention.
      Some examples include mixtures of skew-normal distributions \citep{Lin2007norm}, mixtures of skew-$t$
      distributions \citep{Lin2007t,Pyne2009,Lin2010,vrbik2012,murray2014}, mixtures of generalized hyperbolic
      distributions \citep{MGHD,wei2019mixtures}, mixtures of variance-gamma distributions \citep{mcnicholas2017mixture},
      and mixtures of multivariate normal inverse Gaussian (MNIG) distributions \citep{Karlis2009,Subedi2014,o2016clustering}.\par

        The MNIG distribution \cite{BarndroffNielsen1997} is a  mean-variance mixture of the multivariate normal distribution with the inverse Gaussian distribution. Mixtures of MNIG distributions were first proposed by
    \cite{Karlis2009} and parameter estimation was done using an expectation-maximization (EM) algorithm. Problems
    with the EM algorithm for mixture model
estimation can include slow convergence and unreliable results
arising from an unpleasant likelihood surface. These are
well-known and have been discussed by \cite{titterington1985},
among others. \cite{Subedi2014} implemented an alternative
variational Bayes framework for parameter estimation for these
MNIG mixtures. Although variational inference tends to be faster
than traditional MCMC based approaches and can be easily scaled
for larger datasets, it is an approximation to the true posterior
and the statistical properties of these estimates are not as well
understood \citep{blei2017}. Moreover, it does not provide exact
coverage as noted by \cite{blei2017}.

Some early work in a Bayesian framework for parameter estimation
for the  mixtures of distribution dates back to over three decades
\citep{diebolt1994,robert1996,richardson1997,stephens2000bayesian,Stephens2000}.
Treating the parameter as a random variable, inference regarding
the parameters is conducted based on their posterior
distributions. Additionally, a Bayesian approach can moderate
failure to convergence that is routinely encountered in an EM
algorithm by smoothing the likelihood \citep{FraleyRaftery2002}.
\cite{fruhwirth2006finite} provides a detailed overview of the
Bayesian framework for modeling finite mixtures of distributions.
Recently, the Bayesian framework for modeling skewed mixtures are
gaining more and more attention in the recent literature.
\cite{Sylvia2010} implemented a Bayesian approach to parameter
estimation and clustering through finite mixtures of univariate
and multivariate skew-normal and skew-$t$ distributions.
\cite{hejblum2019sequential} proposed a sequential Dirichlet
process mixtures of multivariate skew-$t$ distributions for
modeling flow cytometry data. \cite{maleki2019robust} introduced
mixtures of unrestricted skew-normal generalized hyperbolic
distributions.

In this paper, we implement a fully Bayesian approach for
parameter estimation via Gibbs sampling.  The structure of the
paper is as follows. In Section \ref{model}, MNIG distributions
and a $G$-component mixtures of them are described. In Section
\ref{parameterestimation}, details on the Gibbs sampling-based
approach to parameter estimation, convergence diagnostics and
label-switching issues are discussed. Here, we also propose novel
approaches to simulate from two distributions: generalized inverse
Gaussian distribution (GIG) and matrix generalized inverse
Gaussian (MGIG) distributions. In section \ref{results}, through
simulation studies and real data analysis, we demonstrate
competitive clustering results. Lastly, Section \ref{discussion}
concludes with a discussion and some future directions.

   \section{Mixtures of Multivariate Normal-inverse Gaussian Distributions}
\label{model}
    The MNIG distribution is based on a mean-variance mixture of a $d$-dimensional multivariate normal with an inverse Gaussian distribution \citep{BarndroffNielsen1997}. Suppose $\Yv|U = u$ is a random vector from a $d$-dimensional multivariate normal distribution with mean $\mu+u\Deltav\beta$ and covariance $u\Deltav$, and $U$ arises from an inverse Gaussian distribution with parameters $\gamma$ and $\delta$ 
    such that $$ \Yv|U = u \sim N (\mu+u\Deltav\beta,u\Deltav) \quad \text{and}\quad U\sim IG(\gamma,\delta).$$ The
    probability density of $U$ is given by
    \begin{equation*}
    f(u) = \frac{\delta}{\sqrt{2\pi}}\exp(\delta\gamma)u^{-3/2}\exp\left\{-\half\left(\frac{\delta^2}{u}+\gamma^2u\right)\right\}.
    \end{equation*}
    The marginal distribution of $\Yv$ is a MNIG distribution with density
    \begin{equation*}
    f_\Yv(\yv) = \frac{\delta}{2^{\frac{d-1}{2}}}\left[\frac{\alpha}{\pi q(\yv)}\right]^{\frac{d+1}{2}}\exp\left({p(\yv)}\right)~K_{\frac{d+1}{2}}(\alpha q(\yv)),
    \end{equation*}
    where \[\alpha = \sqrt{\gamma^2 + \betav^\top \Deltav\betav},\quad p(\yv) = \delta\gamma + \betav^\top(\yv - \muv),\quad q(\yv) = \sqrt{\delta^2 + (\yv-\muv)^\top \Deltav^{-1}(\yv-\muv)},\]
    and $K_{\frac{d+1}{2}}$ is the modified Bessel function of the third kind of order $\frac{d+1}{2}$. For identifiability reasons, $|\Deltav| = 1$ is required. \par
    Combining the conditional $d$-dimensional multivariate normal density   of $\Yv|U=u$ with the marginal density of $U$, the joint probability density can be derived:
    \begin{equation*}
    \begin{split}
    f(\yv,u) = &f(\yv|u)f(u)\\
    = &(2\pi)^{-1/2}|u\Deltav|^{-1/2}\exp\left\{-\half (\yv - \muv - u\Deltav\betav)(u\Deltav)^{-1}(\yv - \muv - u\Deltav\betav)^\top \right\}\\
    \times & \frac{\delta}{\sqrt{2\pi}}\exp(\delta\gamma)u^{-3/2}\exp\left(-\half\left(\frac{\delta^2}{u}+\gamma^2u\right)\right)\\
    \propto &~ u^{-\frac{d+3}{2}}|\Deltav|^{-1/2}\delta\\
    \times & \exp\left\{\delta\gamma -\half\left(\frac{\delta^2}{u}+\gamma^2u\right) -\half (\yv - \muv - u\Deltav\betav)(u\Deltav)^{-1}(\yv - \muv - u\Deltav\betav)^\top\right\}\\
    \end{split}
    \end{equation*}
    By letting $\boldsymbol{\theta} = (\muv, \betav, \delta,\gamma, \Deltav)$, it is easy to show that the above joint density is
    within the exponential family, i.e.,
    \begin{equation*}
    f(\yv,u|\boldsymbol{\theta}) \propto h(\yv,u)r(\boldsymbol{\theta})\exp \left\{ Tr\left(\sum_{j = 1}^{5}\phi_j(\boldsymbol{\theta})t_j(\yv,u)\right) \right\},
    \end{equation*}
    where
    \begin{align*}
    h(\yv,u) &= u^{-\frac{d+3}{2}}, &r(\boldsymbol{\theta}) &= \delta\exp{(\delta\gamma)}\exp\left\{-Tr(\betav\muv^\top)\right\};\\
    \phi_1 &= \betav, &\tv_1(\yv,u) &= \yv^\top;\\
    \phi_2 &= \muv^\top \Deltav^{-1}, &\tv_2(\yv,u) &= \yv^\top u^{-1};\\
    \phi_3 &= -(\Deltav\betav\betav^\top + \frac{\gamma^2}{d}\Id_{d}), &t_3(\yv,u) &= \half u;\\
    \phi_4 &= -(\Deltav^{-1}\muv\muv^\top + \frac{\delta^2}{d}\Id_{d}), &t_4(\yv,u) &= \half u^{-1};\\
    \phi_5 &= -\Deltav^{-1}, &t_5 &= \half\yv\yv^\top u^{-1}.
    \end{align*}

         A $G$-component finite mixture of multivariate normal inverse Gaussian density can be written as
    \begin{equation*}
        f(\mathbf{y}\mid \boldsymbol{\vartheta})= \sum_{g=1}^G \pi_g f(\mathbf{y}\mid\boldsymbol{\theta}_g),
    \end{equation*}
    where $G$ is the number of components, $f(\yv\mid\boldsymbol{\theta}_g)$ is the
    $g^{th}$
      component MNIG density with parameters
    $\boldsymbol{\theta}_g= (\muv_g, \betav_g, \delta_g,\gamma_g, \Deltav_g)$ and $\pi_g>0$ is the mixing proportion
    ($\sum_{g=1}^G\pi_g=1$), and $\boldsymbol{\vartheta}=(\pi_1,\ldots,\pi_G,\boldsymbol{\theta}_1,\ldots,\boldsymbol{\theta}_G)$.
     Consider $n$ independent observations $\yv = (\yv_1,\dots,\yv_n)$ coming from a $G$-component mixture of MNIG distributions. The likelihood for a $G$-component mixture of MNIG distributions can be written as
        \begin{equation*}
    \Lc(\boldsymbol{\theta}_1,\ldots,\boldsymbol{\theta}_G) = \prod_{i=1}^{n}\sum_{g=1}^G \pi_g f_g(\yv_i\mid\muv_g, \betav_g, \delta_g,\gamma_g, \Deltav_g).
    \end{equation*}
     A component indicator variable $\mathbf{Z}_i = (Z_{i1},Z_{i2},\ldots,Z_{iG})$ can be defined such that
     \begin{equation*}
z_{ig}= \left\{
\begin{array}{rl}
1 & \text{if the subject $i$ belongs to group $g$},\\
0 & \text{otherwise.}
\end{array} \right.
\end{equation*}
    which is treated as missing data. The complete-data (i.e., observed data $\yv_i$, the latent variable $u_{ig}$, and the missing data $z_{ig}$) likelihood of the mixture model is given by
    \begin{equation*}
    \begin{split}
    \Lc_{c}(\boldsymbol{\theta}_1,\dots,\boldsymbol{\theta}_G) = &\prod_{g = 1}^{G}\prod_{i=1}^{n}[\pi_g f(\yv_i,u_{ig}|\muv_g, \betav_g, \delta_g, \gamma_g, \Deltav_g)]^{z_{ig}}\\
    = &\prod_{g = 1}^{G}\left\{[r(\boldsymbol{\theta}_g)]^{t_{0g}}\cdot\prod_{i = 1}^{n}[h(\yv_i,u_{ig})]^{z_{ig}}\times\exp Tr\left\{\sum_{j = 1}^{5}
    \phi_{jg}(\boldsymbol{\theta}_g)t_{jg}(\yv,\uv_{g})\right\}\right\},\\
    \end{split}
    \end{equation*}
    where $t_{0g} = \sumn{z_{ig}}$. Also, $\boldsymbol{\theta}_g = (\pi_g, \muv_g, \betav_g, \delta_g,\gamma_g, \Deltav_g)$ denote the parameters related to the $g^{th}$ component.
    Accordingly, the component-specific functions for the parameters, $\phi_{jg}$, and the sufficient statistics $t_{jg}$, for $j = 1,\dots,5$, are given as follows:
    \begin{align*}
    \phi_{1g} &= \betav_g, &\tv_{1g}(\yv,\uv_{g}) &= \sumn z_{ig}\yv_i^{\top};\\
    \phi_{2g} &= \muv_g^\top \Deltav_g^{-1}, &\tv_{2g}(\yv,\uv_{g}) &= \sumn z_{ig}u_{ig}^{-1}\yv_{i}^\top;\\
    \phi_{3g} &= -(\Deltav_g\betav_g\betav_g^\top + \frac{\gamma_g^2}{d}\Id_{d}), &t_{3g}(\yv,\uv_{g}) &= \half \sumn z_{ig}u_{ig};\\
    \phi_{4g} &= -(\Deltav_g^{-1}\muv_g\muv_g^\top + \frac{\delta_g^2}{d}\Id_{d}), &t_{4g}(\yv,\uv_{g}) &= \half \sumn z_{ig}u_{ig}^{-1};\\
    \phi_{5g} &= -\Deltav_g^{-1}, &\tv_{5g}(\yv,\uv_{g}) &= \half \sumn z_{ig} u_{ig}^{-1}\yv_i\yv_i^\top.
    \end{align*}
To  simplify  notation we will use $\tv_{jg}
=\tv_{jg}(\yv,\uv_{g})$, $j=1,\ldots,5$.

   \section{Parameter Estimation using Gibbs Sampling} \label{parameterestimation}

We denote as $GIG(\lambda,\delta^2, \gamma^2)$ the GIG
distribution with probability density function  given by
\[ f(z;\lambda,\delta,\gamma) = \left( \frac{\gamma}{\delta}
\right)^{\lambda} \frac{z^{\lambda-1}}{2K_{\lambda}(\delta
\gamma)} exp \left( - \frac{1}{2} \left ( \frac{\delta^2}{z}+
\gamma^2z \right) \right). \label{GIG} \] Also we denote as
$MGIG_d(\cdot,\cdot,\cdot)$ a MGIG distribution (see Appendix \ref{mGIGapp} for
details). Obviously if the dimension $d=1$ the MGIG is simply a
GIG distribution.

   \subsection{Prior and Posterior distributions}
    If the conjugate prior of $\boldsymbol{\theta}_g$ is of the form
    \begin{equation*}
        p(\boldsymbol{\theta}_g) \propto [r(\boldsymbol{\theta}_g)]^{a_{g,0}^{(0)}}\cdot\exp Tr\left\{\sum_{j = 1}^{5}\phi_{jg}(\boldsymbol{\theta}_g)\av_{g,j}^{(0)}\right\},
    \end{equation*}
    with hyperparameters $\left\{a_{g,0}^{(0)}, \av_{g,1}^{(0)}, \av_{g,2}^{(0)}, a_{g,3}^{(0)}, a_{g,4}^{(0)}, a_{g,5}^{(0)}\right\}$, then the posterior distribution is of the form
    \begin{equation*}
        p(\boldsymbol{\theta}_g|\cdot) \propto [r(\boldsymbol{\theta}_g)]^{a_{g,0}^{(0)}+t_{0g}}\cdot\exp Tr\left\{\sum_{j = 1}^{5}\phi_{jg}(\boldsymbol{\theta}_g)\left(\av_{g,j}^{(0)}+t_{jg}(\yv,\uv_{g})\right)\right\}.
    \end{equation*}
    Hence, the hyperparameters of the posterior distribution are of the form \[\av_{g,j} = \av_{g,j}^{(0)}+\tv_{jg}, \quad j = 0,1,2,3,4,5\]\par
    The posterior distribution of $U|\Yv = \yv$ is a generalized-inverse Gaussian distribution such that
        \begin{equation*}
        U|\Yv = \yv \sim \GIG\left(\dfrac{d+1}{2},q^2(\yv),\alpha^2\right).
    \end{equation*}\par
    A conjugate Dirichlet prior distribution with hyperparameters $(a_{1,0}^{(0)},\dots,a_{G,0}^{(0)})$ is assigned to the mixing weights $(\pi_1,\dots,\pi_G)$. The resulting posterior is a Dirichlet distribution with hyperparameters $(a_{1,0},\dots,a_{G,0})$.\par
    A conjugate gamma prior with hyperparameters $\left(\dfrac{a_{g,0}^{(0)}}{2}+1, a_{g,4}^{(0)}-\dfrac{{a_{g,0}^{(0)}}^2}{4a_{g,3}^{(0)}}\right)$is assigned to $\delta_g^2$, which yields a posterior distribution
    \begin{equation*}
        \delta_g^2 \sim Gamma \left(\dfrac{a_{g,0}}{2}+1, a_{g,4}-\dfrac{{a_{g,0}}^2}{4a_{g,3}}\right).
    \end{equation*}
    Conditional on $\delta_g$, a truncated Normal conjugate prior is assigned to $\gamma_g$. The resulting posterior distribution is given as:
    \begin{equation*}
        \gamma_g|\delta_g \sim \normal\left(\dfrac{a_{g,0}\delta_g}{2a_{g,3}},\dfrac{1}{2a_{g,3}}\right)\cdot\1v\left(\gamma_g > 0\right)
    \end{equation*}

    A conjugate multivariate normal prior conditional on $\Deltav_g$ was assigned to \((\muv_g,\betav_g)\) and the resulting posterior is a multivariate normal distribution conditional on $\Deltav_g$.
    \begin{equation*}
    \left.\begin{pmatrix}
    \muv_g\\
    \betav_g
    \end{pmatrix}\right\vert\Deltav_g\sim MVN
    \left[\begin{pmatrix}
    \frac{2\av_{g,2}a_{g,3} - a_{g,0}\av_1}{4a_{g,3}a_{g,4} - a_{g,0}^2}\\
    \frac{\Deltav_g^{-1}(2\av_{g,1}a_{g,4} - a_{g,0}\av_{g,2})}{4a_{g,3}a_{g,4} - a_{g,0}^2}
    \end{pmatrix},
    \begin{pmatrix}
    \Sigmav_{\mu_g} & \Sigmav_{\mu_g,\beta_g}\\
    \Sigmav_{\mu_g,\beta_g} & \Sigmav_{\beta_g}
    \end{pmatrix}\right],
    \end{equation*}
    where $\Sigmav_{\mu_g} = \dfrac{2a_{g,3}\Deltav_g}{4a_{g,3}a_{g,4} - a_{g,0}^2}, \Sigmav_{\beta_g} = \dfrac{2a_{g,4}\Deltav^{-1}_g}{4a_{g,3}a_{g,4} - a_{g,0}^2}$ and $\Sigmav_{\mu_g,\beta_g} = \dfrac{-a_{g,0}\Id_d}{4a_{g,3}a_{g,4} - a_{g,0}^2}$.\par
    An inverse-Wishart$(\nu_0, \Lambda_0)$ prior is assigned to $\Deltav_g$ and the resulting posterior distribution is an MGIG distribution:
    \begin{equation*}
        \Deltav_g \sim MGIG_d \left(-\frac{\nu_0+t_{0g}}{2}, \betav_g 2t_{3g}\betav_g^\top, {S_0}_g+\Lambda_0\right),
    \end{equation*}
    where ${S_0}_g = \sumn{z_{ig}u_{ig}^{-1}(\yv_i - \muv_g)(\yv_i - \muv_g)^\top}$.\par
    Given an estimate of $\boldsymbol{\theta}_g$, i.e. $(\hat\pi_g, \hat\muv_g, \hat\betav_g, \hat\gamma_g, \hat\delta_g, \hat\Deltav_g)$, the probability that $z_{ig} = 1$ is
    \begin{equation*}
        \widehat{z_{ig}} = \frac{\hat\pi_g f_g(\yv_i|u_{ig};\hat\muv_g,\hat\betav_g,
            \hat\Deltav_g)\cdot f_g(u_{ig};\hat\gamma_g)}{\sum_{k = 1}^{G}{\hat\pi_k f_k(\yv_i|u_{ik};\hat\muv_k,\hat\betav_k,
                \hat\Deltav_k)\cdot f_k(u_{ik};\hat\gamma_k)}}.
    \end{equation*}
    Further details are provided in Appendix \ref{math_detail}.

    \subsection{Gibbs Sampling}
     Parameter estimation in a Bayesian framework can be done by using samples from the posterior  distribution via Gibbs sampling.
     Gibbs sampling is a type of Monte Carlo Markov Chain sampling algorithm which iterates between the following two steps:
     drawing sample of $\zv_i = (z_{i1},\dots,z_{iG})$ from its multinomial conditional posterior distribution with
    \[p(z_{ig}=1|\boldsymbol{\vartheta}^{(t-1)},\yv),\] where
    $\boldsymbol{\vartheta}^{(t-1)} = (\pi_1,\dots,\pi_G)^{(t-1),\thetav_1,\dots,\thetav_G}$
    are the values of $\boldsymbol{\vartheta}$ after $(t-1)-$th iteration, and $g = 1,\dots,G$; and updating
    $\thetav_1,\dots,\thetav_G$ by drawing samples from their posterior distributions and updating $\pi_1,\dots,\pi_G$.\par

    A Gibbs sampling framework for parameter estimation of the MNIG mixtures is summarized as follows:
    \begin{enumerate}
        \item[Step 0] Initialization: For the observed data $\yv = (\yv_1,\dots,\yv_n)$, the algorithm is
         initialized with $G$ components. $\hat z_{ig}$ is initialized using the result from $k$-means clustering (or any other clustering algorithm).
          Based on the initialized $\hat z_{ig}$, parameters from the $g$-th component are initialized as follows:
        \begin{enumerate}
            \item $\gamma_g=\delta_g =1$.
            \item $\muv_g$ is set as the component sample mean.
            \item $\betav_g$ is assigned a $d$-dimensional vector with all entries equal to $0.05$.
            \item $\Deltav_g$ is initialized as the component sample variance
            matrix divided by its determinant rise to power $1/d$ to ensure that  $\left\vert \Deltav_g\right\vert = 1$.
            \item $\pi_g$ is set as the proportion of observation in the $g^{th}$ component.
        \end{enumerate}
        \item[Step 1] At $t^{th}$ iteration, update $z_{ig}$ from its multinomial conditional posterior distribution with \[\Pr(z_{ig}=1) = \hat z_{ig} = \frac{\pi_g f(\yv_i,u_{ig}|(\muv_g, \betav_g, \delta_g, \gamma_g, \Deltav_g)^{(t-1)})}{\sum_{k=1}^{G}\pi_k f(\yv_i,u_{ik}|(\muv_k, \betav_k, \delta_g, \gamma_k, \Deltav_g)^{(t-1)})}.\]
        where $(\muv_g, \betav_g, \delta_g, \gamma_g, \Deltav_g)^{(t-1)}$ are the values of $\muv_g, \betav_g, \delta_g, \gamma_g$, and $\Deltav_g$ sampled from the $t-1^{th}$ iteration.
        \item[Step 2]
        \begin{enumerate}
            \item Update $U_i$'s by drawing samples from $\GIG((d+1)/2,q^2(\yv_i),\alpha^2)$ distribution for $i = 1,\dots, n$;
            \item Based on the updated $U_{ig}$ and $z_{ig}$, update the hyperparameters $\{a_{g,0},\av_{g,1},\av_{g,2},a_{g,3},a_{g,4},a_{g,5}\}$, $g = 1,\dots, G$;
            \item Update the parameters to  $(\muv_g,\betav_g,\delta_g,\gamma_g,\Deltav_g)^{(t)}$ by each drawing one sample from their posterior
            distributions described in the previous section;
            \item Update $(\pi_1^{(t)},\dots,\pi_G^{(t)})$ from $ \text{Dir}(a_{1,0},\dots,a_{G,0})$.
        \end{enumerate}
    \end{enumerate}
    Step 1 and 2 are iterated until convergence.\par

    \subsection{Sampling from GIG and MGIG Distributions}
    Note that posterior distribution of $U|\Yv = \yv$ is a GIG  distribution and the conditional posterior distribution
    of $\Deltav_g$ is a MGIG distribution. Hence we need to simulate from them. While for the GIG the literature has certain approaches,
    here, we propose a novel approach to sample from the univariate
    GIG distribution and the  MGIG distribution suitable for the
    MCMC approach to be used.

  \subsubsection{Sampling from Univariate GIG Distributions}

    At the past there are attempts for simulating from GIG based on
different approaches
\citep{atkinson1982simulation,Dagpunar1989,devroye2014random,lai2009generating,hormann2014generating,leydold2011generating}.
Our approach is particularly suitable as it is based on a Markov
chain and hence it can be embedded in the MCMC easily.

    The approach for sampling from a univariate GIG distribution is based on a property of the GIG distribution. Firstly,
    if $X$ is a GIG random variable, then the same is true for $X^{-1}$. Secondly, if $X$ and $Y$ are two independent random variables,
     from a GIG distribution and a Gamma distribution, respectively, then the transformed random variables $(X+Y)^{-1}$ and $X^{-1} - (X+Y)^{-1}$
     follow a GIG distribution and a Gamma distribution separately, and they are independent \citep{Letac1983, Letac2000}.
     It has been further shown by \cite{Letac1983} that, if $X$ is a GIG random variable, and $Y$ and $Z$ are Gamma random variables,
     then the following holds:
    \begin{equation*}
    X \stackrel{d}{=} Y + \frac{1}{Z+\frac{1}{X}},
    \end{equation*}
    where $\stackrel{d}{=}$ indicates that the left hand side and the right hand side possess the same distribution.\par
    The fraction form the right hand side of above is a random continued fraction, denoted as $[Y;Z,X]$. Following the result of \cite{Goswami2004}, consider the Markov Chain defined recursively by continued random fractions $X_1 = [Y_1;Z_1,X_0]$, $X_2 = [Y_2;Z_2,X_1]$ and in general $X_n = [Y_n;Z_n,X_{n-1}]$, they must converge to a stationary distribution. In details, if $\{Y_i,Z_i, i = 1,2,\dots\}$ are independent random variables from $Gamma(\lambda, \alpha)$ and $Gamma(\lambda,\beta)$ distributions respectively, then the chain converges to the density of a $\GIG(\lambda,2\alpha,2\beta)$ distribution.\par
    The algorithm to sample random numbers $U$ from a $\GIG\left(-\frac{d+1}{2}, q^2(\yv), \alpha^2,\right)$ distribution is outlined below:
    \begin{itemize}
        \item[Step 1]: Start with a $U$ from any distribution.
       \item[Step 2]: Generate $$V \sim Gamma\left(-\dfrac{d+1}{2},\dfrac{\alpha^2}{2}\right) ~\text{and}~ S \sim Gamma\left(-\dfrac{d+1}{2},\dfrac{q^2(\yv)}{2}\right).$$
        \item[Step 3]: Update $U$ using $U = S + \dfrac{1}{V+\dfrac{1}{U}}$.
        \item[Step 4] Repeat Step 2 and Step 3 for a number of times until converge.
    \end{itemize}
    Convergence of the above Markov Chain is monitored using the Heidelberger and Welch'™s convergence diagnostic for only one chain.\par

    \subsubsection{Sampling from Matrix GIG distributions}

    Sampling from MGIG is not developed so far to our knowledge.
    In \cite{Fazayeli2016} a standard importance sampling approach is
    described. It is interesting that MGIG is conjugate for the
    covariance matrix of a multivariate normal distribution.

    Our algorithm for sampling from a MGIG distribution is based on the Matsumoto-Yor (MY) property for random matrices of different dimensions, discussed by \cite{Massam2006}.

Let $\mathcal{V}_n$ be the Euclidean space of $n \times n$ real
symmetric matrices. Let also $\mathcal{V}_n^+$ denote the cone of
positive definite matrices in $\mathcal{V}_n$.
    Let $a \in \mathcal{V^+}_s$ and $b \in \mathcal{V^+}_r$, $X$ and $Y$ be two independent random matrices with $MGIG_r$ and Wishart distributions, respectively, with  $p = q+\dfrac{r-s}{2}$:
    \begin{equation*}
    X \sim MGIG_r\left(-p,P(z^\top)a,b \right) \quad \text{and}\quad Y\sim W_s(q,a).
    \end{equation*}
    Let $z$ be a given constant $s\times r$ matrix of full rank, and $P(z^\top)a = z^\top az$
     defining a linear mapping, then, $U$ and $V$ defined as
    \begin{equation*}
    U = (P(z)X+Y)^{-1},\quad V= X^{-1}-P(z^\top)(P(z)X+Y)^{-1}
    \end{equation*}
    are independent $MGIG$ and Wishart distributions and in
    particular
    \begin{equation*}
    U \sim MGIG_s(-q, P(z)b,a)\quad \text{and} \quad V\sim W_r(p,b).
    \end{equation*}
    The detailed simulation of $\Deltav_g\sim MGIG_d\left(-\frac{\nu_0+t_{0g}}{2}, \betav_g 2t_{3g}\betav_g^\top, {S_0}_g+\Lambda_0\right)$ is organized as:\par
    Suppose we take $r = 1, s = d, q = \dfrac{\nu_0+t_{0g}}{2}$ and $p = q+\dfrac{1-d}{2}$; let $z = \betav_g$, $a = \Lambda_0+S_{0g}$, where
    ${S_0}_g = \sumn{z_{ig}u_{ig}^{-1}(\yv_i - \muv_g)(\yv_i - \muv_g)^\top}$ and let $b = 2t_{3g}$. Then, take the following steps:
    \begin{itemize}
        \item[Step 1] Simulate an $X$ following the univariate generalized inverse Gaussian distribution such that
        \[X \sim GIG \left(-p,z^\top az,b \right).\]
        \item[Step 2] Simulate a $d\times d$ matrix variable $W$ following Wishart distribution, i.e.,
        \[W \sim W_d\left(q,a\right).\]
        \item[Step 3] Compute $Y = (zXz^\top + W)^{-1}$; it follows a $MGIG_{d}(-q,zbz^\top,a)$ distribution.
    \end{itemize}

The algorithm just described offers a simple way to simulate from
the MGIG distribution.  In order to generate from a MGIG we need
just to simulate from a univariate GIG distribution and a Wishart
distribution which is very standard. Also, note that with the
algorithm above GIG is simulated  as a Markov chain and this is
very suitable in the Gibbs setting we are discussing.

    \subsection{Convergence Assessment and Label Switching Issue}
    In the Bayesian approach to parameter estimation and clustering in finite mixture models, label
switching can occur. This is well known and refers to the
likelihood not changing when the mixture components  are permuted
\citep{Stephens2000}. There are many different approaches to
dealing with the label switching issue in the literature. For
example, \cite{richardson1997} suggested that one can put
artificial constraints on the mixing proportion
$\pi_1,\dots,\pi_G$ or the parameters to force the labeling to be
unique. \cite{Celeux2000} considered a decision theoretic approach
to overcome the label switching issue. \cite{Stephens2000}
proposed an algorithm that combined the relabeling algorithm with
the decision theoretic approach. In our analysis, we found that
relabeling the parameter estimates by the constraint on the mixing
proportion works well.\par
    To diagnose the convergence of Monte Carlo Markov Chains, three independent sequences, with different $k$-means initializations are simulated. Likelihood is calculated using the updated parameters at the end of each iteration and the chains of likelihood monitored for convergence using the Gelman-Rubin diagnostics \citep{gelman1992} which is based on the comparisons of the between and within variations. As early iterations reflect the starting approximation and may not represent the target posterior, samples from the early iterations known as ``burn-in'' period are discarded \citep{BDA3}. If the potential scale reduction factor calculated based on the likelihood chains after ``burn-in'' is below 1.1, the chains are considered as converged and mixing well. After reaching a stationary approximation of the posterior distribution, averages of the samples, drawn from the approximated posterior in the Markov Chains, discarding those from the ``burn-in'' period, provides good estimation
 s to the parameters \citep{diebolt1994}.\par
    \subsection{Model Selection and Performance Assessment}
Model selection for selecting the optimal number of components can be done by a comparison between models containing different number of clusters \citep{FraleyRaftery2002,BDA3}. Several model selection criteria has been proposed in the literature such as Akaike Information Criterion \citep[AIC;][]{AIC}, Bayesian Information Criterion \citep[BIC;][]{BIC}, Deviance Information Criterion \citep[DIC;][]{DIC}, and Integrated Complete Likelihood \citep[ICL;][]{ICL}. Model selection based on BIC has been shown to obtain satisfactory performance \cite{Leroux1992,Keribin2000,RoederWasserman1997}. Additionally, \cite{SteelRaftery2009} investigated the performance of several model selection approaches including some Bayesian approaches as well as information criteria based approaches for Gaussian mixture models and showed that the BIC performed the best among the ones compared. Hence, here, the Gibbs sampling algorithm is carried out for a range of number of components, $G$ with model selection
  conducted using the BIC.  When the true class label is known, the performance of the clustering algorithm can be assessed using the adjusted Rand index \cite[ARI;][]{hubert1985}. The ARI is a measure of the pairwise agreement between the true and estimated classifications after adjusting for agreement by chance such that a value of `1' indicates a perfect agreement and the expected value ARI  under random classification is 0.

   \section{Simulation Studies and Real Data Analysis} \label{results}
%   \subsection{Simulation Studies}
    \subsection{Simulation Study 1}

\begin{figure}[h!]
        \begin{minipage}{.5\textwidth}
            \centering
            \includegraphics[width=.95\linewidth]{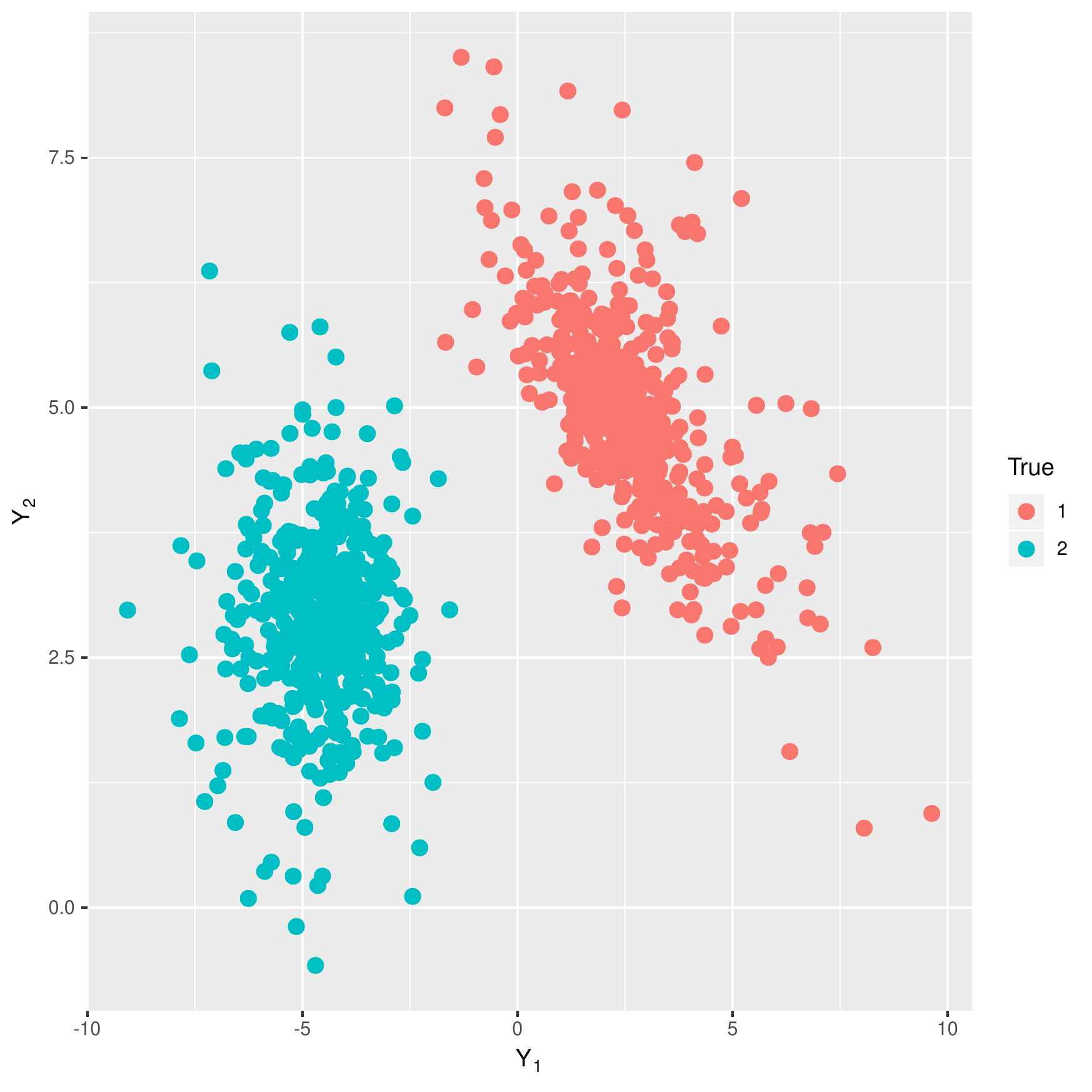}
            %\label{fig:scatter}
        \end{minipage}%
        \begin{minipage}{.5\textwidth}
            \centering
            \includegraphics[width=.95\linewidth]{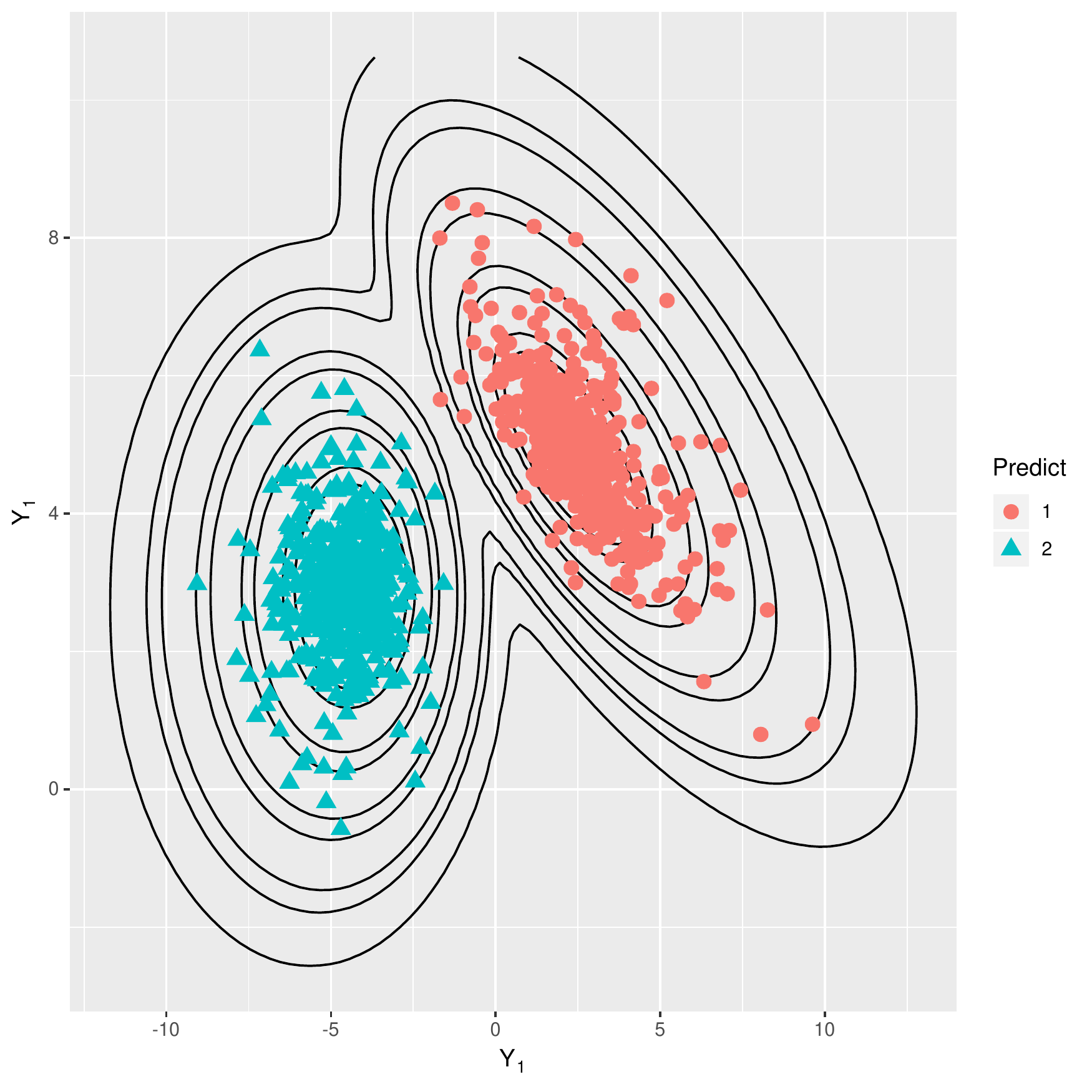}
            %\label{fig:contour}
        \end{minipage}
        \caption{Scatter plot highlighting the true labels (left) and a contour plot showing the predicted classifications (right). Here, the ARI is 1.}
        \label{fig:sim1}
    \end{figure}

        In this simulation study, the proposed algorithm is applied to 100 two-dimensional  data sets with two skewed and heavy-tailed components
        (for both sample size is 500); see Figure~\ref{fig:sim1} (left panel) for one example. The true parameters used to generate the data are summarized in Table~\ref{tab:s1}. The proposed algorithm is run with the number of components ranging from $G = 1$ to $G = 5$. A two component model is selected by the BIC for all 100 datasets with an average ARI is $1.00$ and standard deviation (sd) of $0.00$. The estimated parameters are also summarized in Table~\ref{tab:s1}.\par

    \begin{table}[h!]
        \centering
        \caption{True and Estimated Values for the MNIG Parameters in Simulation Study 1}\label{tab:s1}
        \vspace*{0.1 in}
        \scalebox{0.85}{
        \renewcommand{\arraystretch}{1.3}
        \begin{tabular}{|c|c|c|c|c|}
            \hline  &\multicolumn{2}{|c|}{\textbf{Component 1}}&\multicolumn{2}{|c|}{\textbf{Component 2}}\\
            \hline  & \textbf{True Parameters} & \textbf{Estimated Parameters} & \textbf{True Parameters} & \textbf{Estimated Parameters}\\
                 &  & \textbf{(Standard Error)}&  & \textbf{(Standard Error)}\\
            \hline  $\gamma$ & $0.9$ & $0.97 (0.14)$ & $1.2$ & $1.38 (0.19)$  \\
            \hline $\delta$ & $0.9$ & $0.95 (0.09)$ & $1.2$ & $1.33 (0.15)$ \\
            \hline  $\muv$ & $[2,\quad 5]$ &$[1.96 (0.09),\quad 4.99 (0.07)]$ & $[-4,\quad 3]$ & $[-3.90 (0.13),\quad 3.00 (0.10)]$ \\
            \hline  $\betav$ & $[0.5,\quad 0.5]$ & $[0.56 (0.10),\quad 0.56 (0.13)]$ & $[-0.5,\quad -0.1]$ & $[-0.75 (0.19),\quad -0.14 (0.14)]$\\
            \hline   $\Deltav$ &$\begin{bmatrix} 2&-1\\-1&1\end{bmatrix}$ & $\begin{bmatrix}
            1.98 (0.11) & -0.99 (0.07)\\-0.99 (0.07)&1.00 (0.06)\end{bmatrix}$ &  $\begin{bmatrix} 1 & 0 \\ 0& 1\end{bmatrix}$ & $\begin{bmatrix}
            1.00(0.06) & -0.01 (0.05) \\ -0.01 (0.05) &1.00 (0.06)\end{bmatrix}$ \\\hline
        \end{tabular}}
    \end{table}
    Figure~\ref{fig:sim1} (left) shows the true component membership of one of the hundred datasets and Figure~\ref{fig:sim1} (right) gives the contour plot based on the estimated parameters for the dataset described in the left panel.
Although flexible clustering models based on skewed data have
emerged in the last decade  or so, Gaussian mixture still remain
one of the predominantly used approach for model-based clustering
due to its mathematical tractability and the relative
computational simplicity with parameter estimation. Therefore, the
Gaussian mixture models (GMM) implemented in the {\sf R} package
\texttt{mclust} \citep{mclust} are also applied to these datasets.
However, these Gaussian mixture models cannot account for
skewness. Hence, mixtures of generalized hyperbolic distributions
\citep[MixGHD;][]{MGHD} implemented in the {\sf R} package
\texttt{MixGHD} \citep{MixGHD} are also applied to these datasets.
Mixtures of generalized hyperbolic distributions
\citep{mcneil2015} also have the flexibility of modeling skewed as
well as symmetric components. Several skewed distributions such as
the skew-$t$ distribution, skew normal distribution,
variance-gamma distribution, and MNIG distribution as well as
symmetric distributions such as the Gaussian and $t$-distribution
can be obtained as a limiting case of the generalized hyperbolic
distribution \citep{MGHD}. In this simulation, the Gaussian
mixture models always chooses 3 or 4 components since it needs
more than one component to fit the skew clusters. The mixture of
generalized hyperbolic distributions always choose the
two-component model as well and gives ARI $=1.00$ with a standard
deviation (sd) of $0.00$.\par

    The 95\% credible intervals for all parameter estimation for all hundred datasets are given in the Appendix \ref{supp_fig}, where the lower and upper endpoints of the credible intervals are the empirical 0.025-percentiles and 0.975-percentiles.

    \subsection{Simulation Study 2}
    In this simulation, 100 four-dimensional datasets were generated with three underlying groups where one component is skewed with 200 observations, the other component is symmetric with 200 observations, and the third component is skewed with 100 observations.  The parameters used to generate the data are summarized in Table~\ref{tab:s2}. The proposed algorithm is applied to these datasets for $G=1$ through $G=7$, and it correctly selected the correct three-component model 99 out of 100 times with an average ARI of 1.00 (sd 0.00).\par
    \begin{table}[h!]
        \centering
        \caption{True and Estimated Values for the Parameters in Simulation Study 2}\label{tab:s2}
%        \vspace*{0.01 in}
        \scalebox{0.65}{
            \renewcommand{\arraystretch}{1.0}
            \begin{tabular}{|c|c|c|}
                \hline  &\multicolumn{2}{|c|}{\textbf{Component 1} ($n=100$)}\\
                \hline  & \textbf{True Parameters} & \textbf{Estimated Parameters (Standard Error)} \\
                \hline  $\gamma$ & $0.6$ & $0.78 (0.19)$  \\
                \hline  $\delta$ & $0.6$ & $0.67 (0.10)$  \\
                \hline  $\muv$ & $[9,-6,-5,9]$ & $[8.95 (0.09),-5,96 (0.09),-4.95 (0.09),8.97 (0.09)]$ \\
                \hline  $\betav$ & $[0,0,-0.5,-0.5$ & $[-0.10 (0.20), 0.06 (0.19), -0.33 (0.21), -0.49 (0.26)]$ \\
                \hline   $\Deltav$ & $\begin{bmatrix} 1&0&0&0\\0&1&0&0\\0&0&1&0\\0&0&0&1\end{bmatrix}$ & $\begin{bmatrix}  1.03 (0.15)& 0.01 (0.12)& -0.03 (0.11)&-0.04 (0.11)\\0.01 (0.12)& 1.05 (0.15)& 0.04 (0.13)& 0.02 (0.11)\\-0.03 (0.11)& 0.04 (0.13)&  1.08 (0.15)& 0.00 (0.13)\\ -0.04 (0.11)& 0.02 (0.11)& 0.00 (0.13)& 0.97 (0.12)\end{bmatrix}$  \\
                \hline
                \hline  &\multicolumn{2}{|c|}{\textbf{Component 2} ($n=200$)}\\
                \hline  & \textbf{True Parameters} & \textbf{Estimated Parameters (Standard Error)} \\
                \hline  $\gamma$ & $0.9$ & $0.94 (0.14)$  \\
                \hline  $\delta$ & $0.9$ & $0.94 (0.09)$  \\
                \hline  $\muv$ & $[5,3,0,-7]$ & $[4.96 (0.14), 2.98 (0.09), 0.00 (0.08), -6.99 (0.13)]$ \\
                \hline  $\betav$ &  $[0.5,0.5,0.5,0.5]$ & $[0.36 (0.18), 0.47 (0.19), 0.51 (0.22), 0.82 (0.43)]$ \\
                \hline   $\Deltav$ & $\begin{bmatrix} 2&0&0&1\\0&1&0&0\\0&0&1&0\\1&0&0&1\end{bmatrix}$ & $\begin{bmatrix} 2.19 (0.25)& 0.09 (0.11)& 0.04 (0.14)&1.02 (0.16)\\0.09 (0.11)& 1.02 (0.11)& 0.03 (0.09)& 0.03 (0.09)\\0.04 (0.14)& 0.03 (0.09)& 1.02 (0.10)& 0.00 (0.09)\\1.02 (0.16) & 0.03 (0.09) & 0.00 (0.09) & 0.95 (0.12)\end{bmatrix}$  \\
                \hline
                \hline  &\multicolumn{2}{|c|}{\textbf{Component 3} ($n=200$)}\\
                \hline  & \textbf{True Parameters} & \textbf{Estimated Parameters (Standard Error)} \\
                \hline  $\gamma$ & $1.2$ & $1.40 (0.21)$ \\
                \hline  $\delta$ & $1.2$ & $1.34 (0.16)$ \\
                \hline  $\muv$ & $[-3,-2,7,3]$ & $[-2.93 (0.13), -1.93 (0.13), 6.87 (0.14), 2.93 (0.14)]$\\
                \hline  $\betav$  & $[0,0,0,0]$ & $[-0.05 (0.16), -0.06 (0.15), 0.09 (0.17), 0.06 (0.18)]$\\
                \hline   $\Deltav$ & $\begin{bmatrix} 1&0&0&0\\0&1&0&0\\0&0&1&0\\0&0&0&1\end{bmatrix}$ & $\begin{bmatrix}1.00 (0.09)&0.02 (0.09)&-0.01 (0.08)&0.00 (0.08)\\ 0.02 (0.09)& 1.02 (0.10)& 0.00 (0.09)& 0.01 (0.08)\\-0.01 (0.08)& 0.00 (0.09)& 1.02 (0.11) &0.00 (0.09)\\  0.00 (0.08)& 0.01 (0.08)& 0.00 (0.09)& 1.01 (0.09) \end{bmatrix}$ \\
                \hline

        \end{tabular}}
    \end{table}

\begin{figure}[h!]
        \begin{minipage}{.5\textwidth}
            \centering
            \includegraphics[width=.95\linewidth,height=3 in]{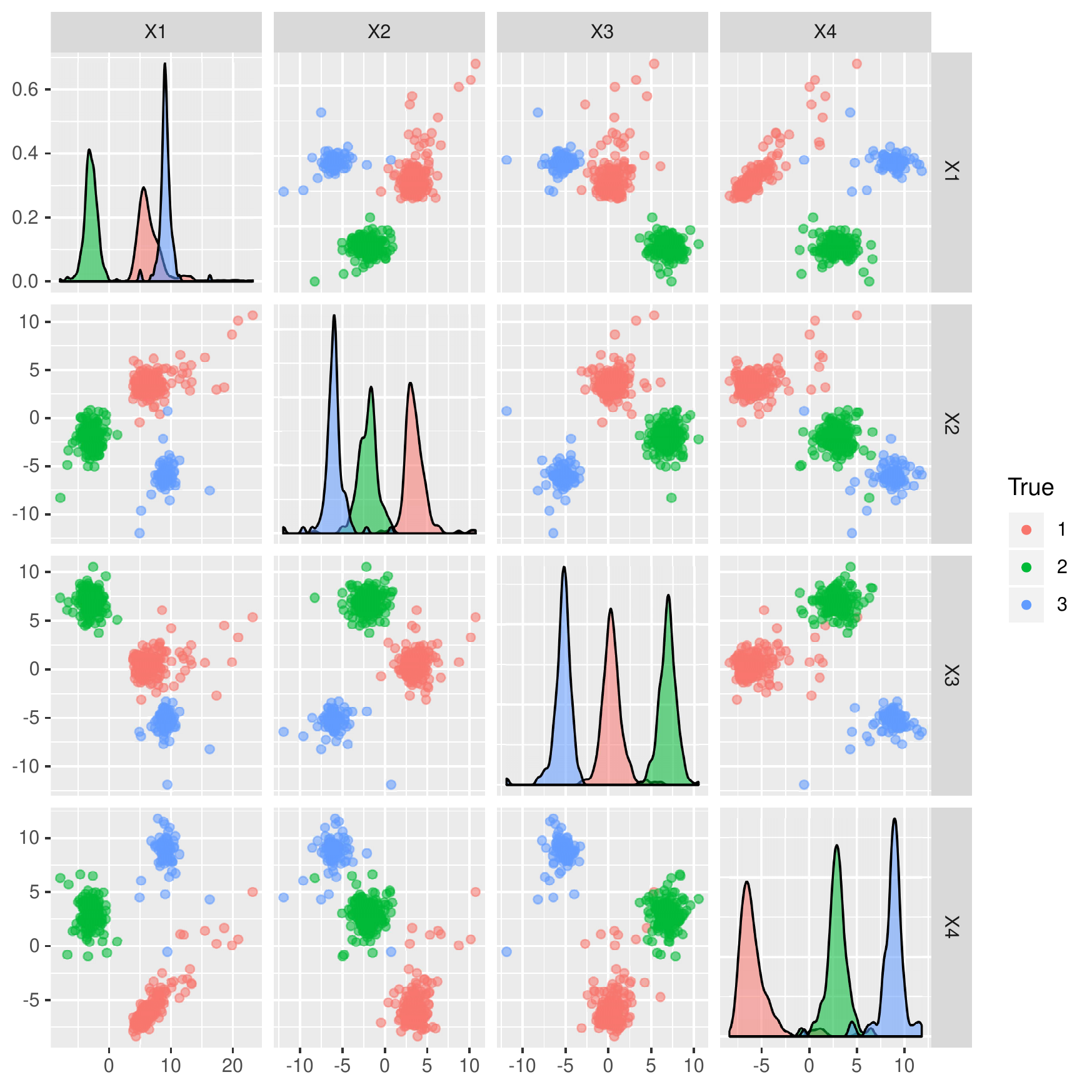}

        \end{minipage}%
        \begin{minipage}{.5\textwidth}
            \centering
            \includegraphics[width=.95\linewidth,height=3 in]{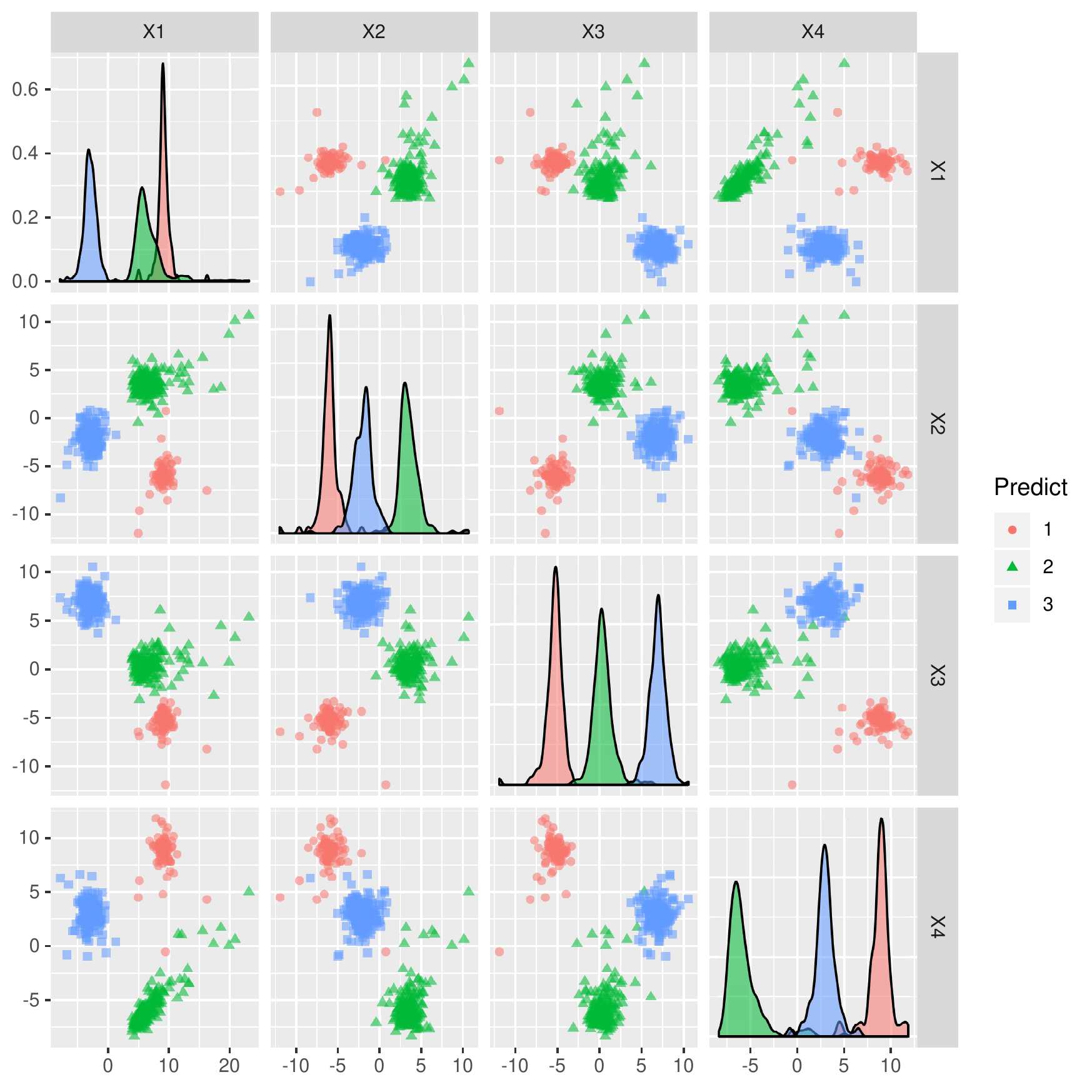}

        \end{minipage}
        \caption{Pairwise scatter plot highlighting the true labels (left) and a pairwise scatter plot showing the predicted classifications (right) of one of the hundred datasets. Here, the ARI was 1.}
        \label{fig:sim2}
    \end{figure}

    The average of the parameter estimates for the 99 datasets provided in Table~\ref{tab:s2} shows good parameter recovery. Figure~\ref{fig:sim2} (right) gives the pairwise scatter plot based on the estimated parameters for this dataset where the true group labels are described in the left panel.\par

    Gaussian mixture models and mixtures of generalized hyperbolic distributions are applied to these datasets. The mixture of generalized hyperbolic distributions correctly selects the three-component model 84 times out of the 100 datasets with an average ARI  of $1.00$ (sd 0.00). The Gaussian mixture models did not choose the correct number of components for all 100 datasets.\par

   \subsection{Real Data Analysis}
    The proposed algorithm is also applied to some benchmark clustering datasets.\\

\begin{figure}[h!]
            \centering
            \includegraphics[scale=0.45]{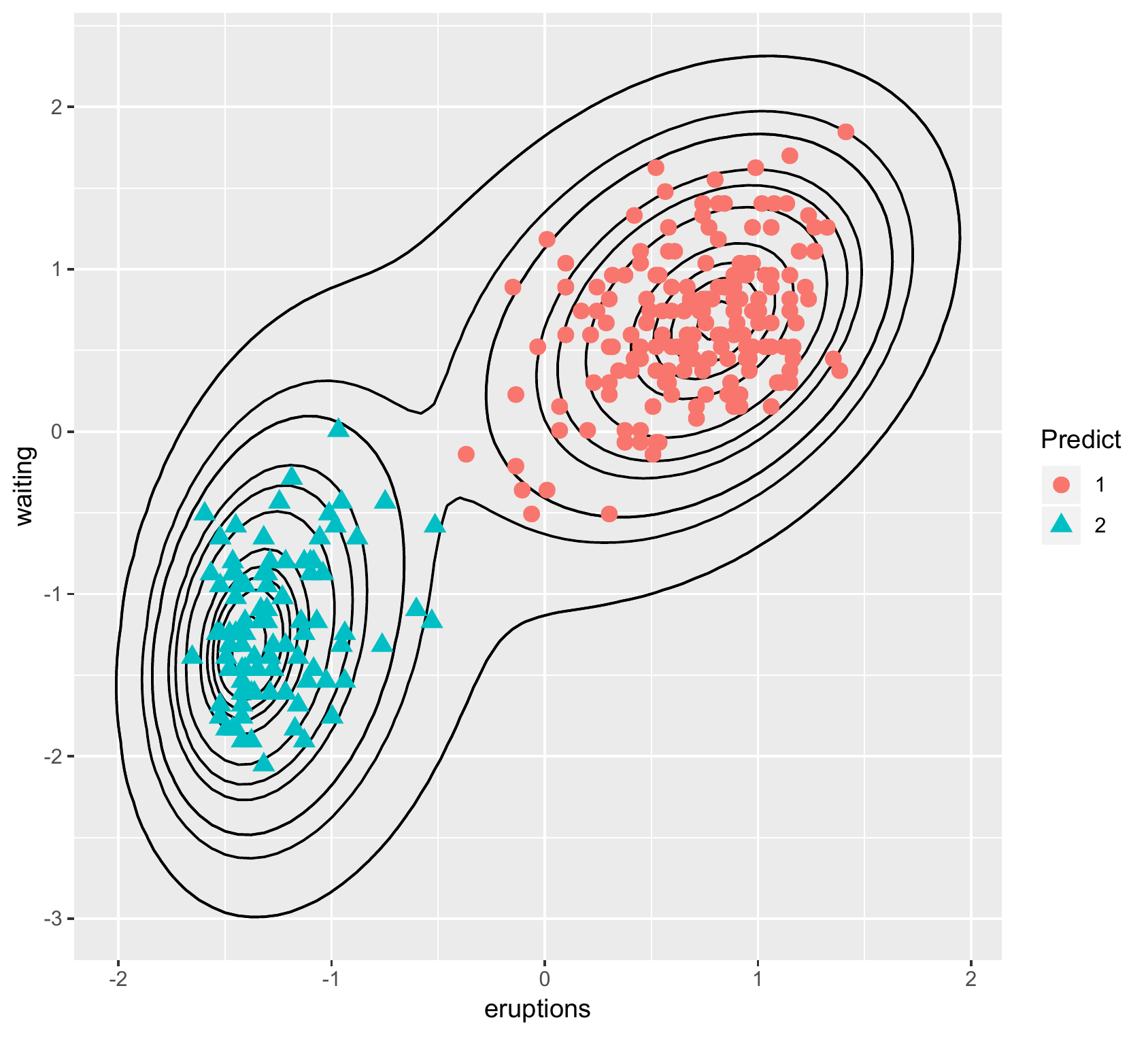}
            \caption{Contour plot for the Old Faithful data using the estimated values of the parameters using the proposed algorithm.}
            \label{fig:contor_faithful}
    \end{figure}

    \noindent \textbf{The Old Faithful Dataset}\\
    The Old Faithful data available in the \texttt{R} package \texttt{MASS} consists of the waiting time between eruptions and the duration of the eruption for the Old Faithful geyser in Yellowstone National Park, Wyoming, USA. There are 272 observations and each observation contains 2 variables (waiting time between eruption and duration of eruption). We ran our algorithm on the scaled data for $G=1~\text{to}~5$. A two component model was selected. This is consistent with \cite{Subedi2014} that used mixtures of MNIG distributions in a variational Bayes framework and with \cite{franczak2014mixtures} and \cite{vrbik2012} who used other skewed mixture models. The contour plot of the fitted model given in Figure~\ref{fig:contor_faithful} shows that our fitted model captures the density of the data fairly well.

\begin{figure}[h!]
        \begin{minipage}{.5\textwidth}
            \centering
            \includegraphics[width=.95\linewidth,height=3 in]{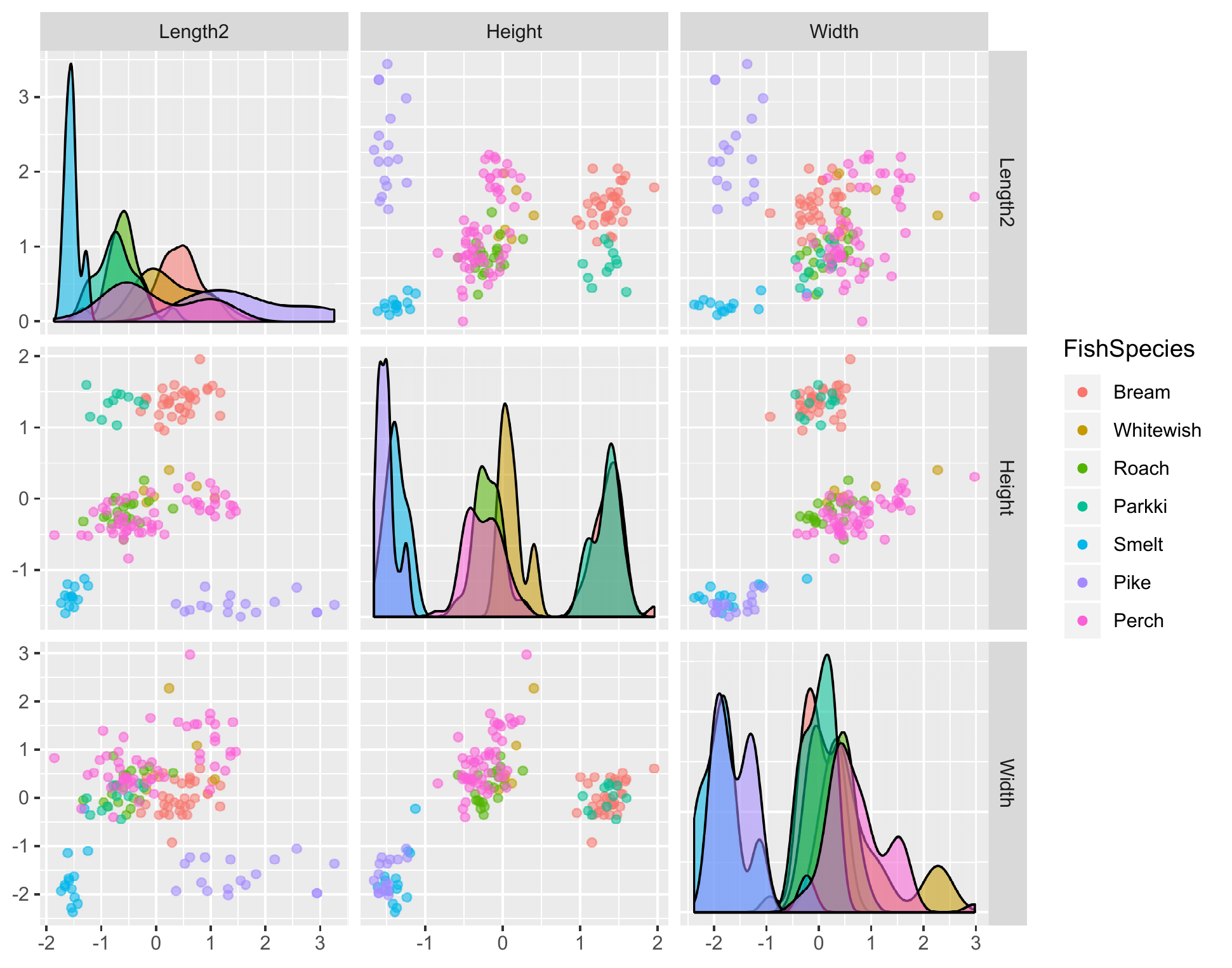}

        \end{minipage}%
        \begin{minipage}{.5\textwidth}
            \centering
            \includegraphics[width=.95\linewidth,height=3 in]{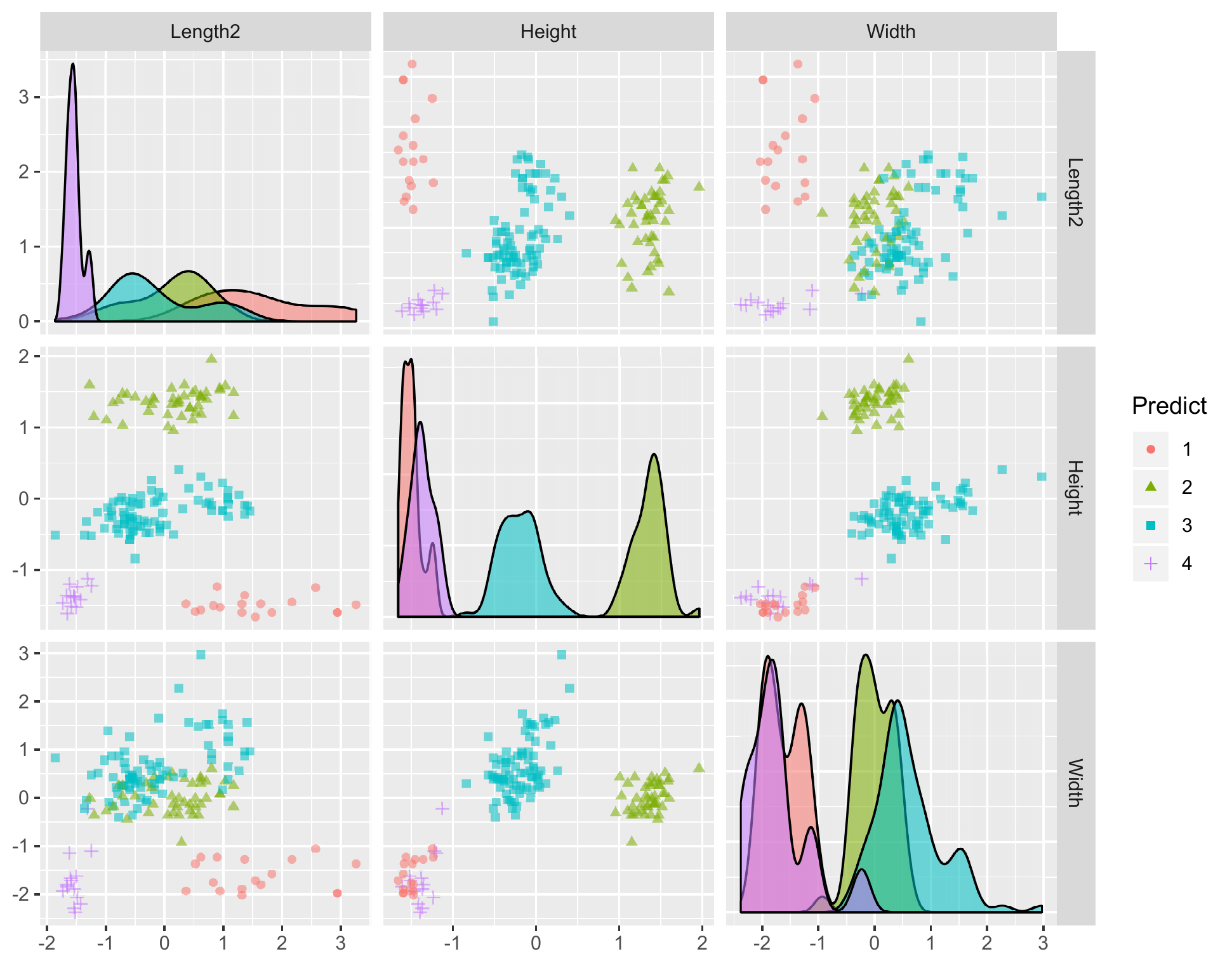}

        \end{minipage}
        \caption{Pairwise scatter plot highlighting the true labels (left) and a pairwise scatter plot showing the predicted classifications (right) for the fish catch data.}
        \label{fig:fishcatch}
    \end{figure}

    \noindent \textbf{The Fish Catch Dataset}\\
    The fish catch data, available from the \texttt{R} package \texttt{rrcov}, consists of the weight and different body lengths measurements of seven different fish species. There are 159 observations in this data set. Similar to \cite{Subedi2014}, after dropping the highly correlated variables, the variables \texttt{Length2}, \texttt{Height} and  \texttt{Width} were used for the analysis where \texttt{Length2} is the length from the nose to the notch of the tail, \texttt{Height} is the maximal height as a percentage of the length from the nose to the end of the tail, and \texttt{Width} is the maximal width as a percentage of the length from the nose to the end of the tail. The proposed algorithm is applied (after scaling the data) for $G=1~\text{to}~9$ and it selects a four-component model. Figure~\ref{fig:fishcatch} shows the pairwise scatter plots for this dataset, with the left panel showing the true species of the fish and the right panel showing the estimated groups. Although 
 the true number of species of fish is seven, from the pairwise scatter plot one can see that the species \texttt{White, Roach} and \texttt{Perch} are hard to indistinguishable and there is no clear separation between the species \texttt{Bream} and \texttt{Parkki}. Table~\ref{tab:fishco} summarizes the cross tabulation of the true species and estimated group membership and it is consistent with the result from \cite{Subedi2014} that used a variational Bayes approach for clustering using mixtures of MNIG. The GMM and mixGHD are also applied to the \texttt{Fish Catch} data and both resulted in a five component model with classification where the additional fifth component contained fish from  both \texttt{Whitewish} and \texttt{Perch} (see Table \ref{tab:fishco} for detail).\par

    \vspace*{0.2 in}

    \begin{table}
            \begin{center}
                    \centering
                    \renewcommand{\arraystretch}{1.3}
                                                \caption{Cross tabulation of true fish species and the estimated classification using the proposed model, Gaussian mixture models, and the mixtures of generalized hyperbolic distributions.}
                    \begin{tabular*}{\textwidth}{c@{\extracolsep{\fill}} cccc|ccccc|ccccc}
                        \hline
                        &\multicolumn{4}{c}{\textbf{Proposed algorithm}}&\multicolumn{5}{c}{\textbf{GMM}}&\multicolumn{4}{c}{\textbf{MixGHD}}&\\
&\multicolumn{4}{c}{ARI: 0.63}&\multicolumn{5}{c}{ARI: 0.52}&\multicolumn{4}{c}{ARI: 0.54}&\\
&\multicolumn{4}{c}{Estimated Groups}&\multicolumn{5}{c}{Estimated Groups}&\multicolumn{4}{c}{Estimated Groups}&\\\hline
                            &   1   &   2   &   3   & 4     &   1   &   2   &   3   & 4 & 5&1   &   2   &   3   & 4 & 5 \\ \hline
                        Bream   &   34  &       &       &   &   34  &       &       & & &   &       &     &34 &\\
                        Parkki  &   11  &       &   & &11   &       &    &  & & &       &   & 11 & \\\hline
                        Whitewish&  &   6   &       &   & &3&3& &3& &       &   3   &    & 3\\
                        Roach & & 20 &  &  & &20& & & & &  & 20 &  &  \\
                        Perch &  & 56 &  & & & 36 & 20 &  & 20&  &  & 36 &  & 20 \\\hline
                        Smelt & &  & 14 &  & & &  & 11 & 3&14 &  &  & & \\\hline
                        Pike &  &  & & 17&  &  & &  & 17& & 17 & &  & \\\hline
                \end{tabular*}
                \label{tab:fishco}
            \end{center}
            \end{table}

\begin{figure}
            \centering
            \includegraphics[scale=0.4]{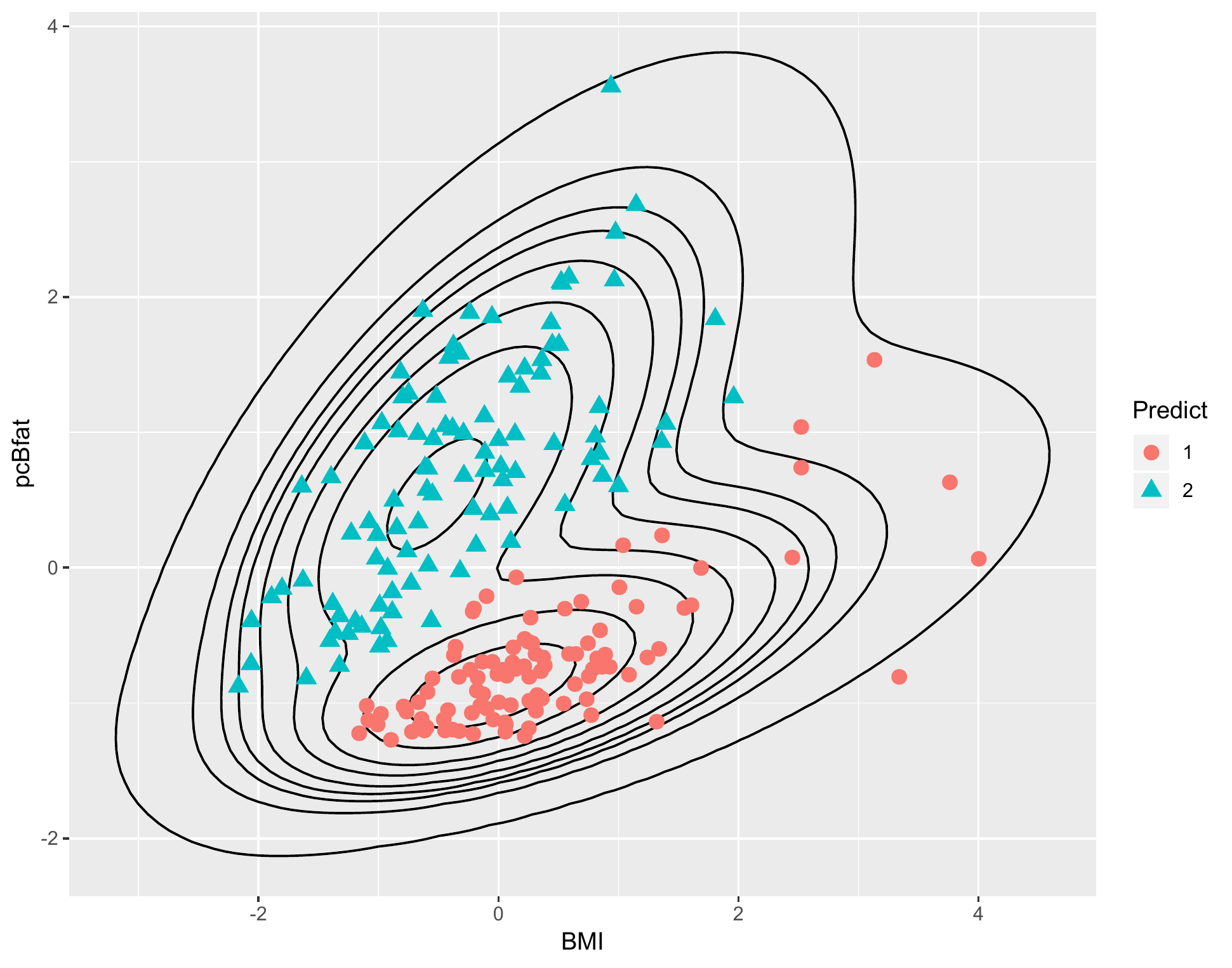}
            \caption{Contour plot using the estimated parameters for the AIS data}
            \label{fig:contor_ais}
        \end{figure}

        \noindent \textbf{The Australian Athletes (AIS) Dataset}:\\
        The AIS dataset available in the \texttt{R} package \texttt{DAAG} \citep{DAAG} contains 202  observations and 13 variables comprising of measurements on various characteristics of the blood, body size, and sex of the athlete. The proposed algorithm is applied on a subset of dataset with the variables body mass index (\texttt{BMI}) and body fat (\texttt{Bfat}) as has been previously used \citep{vrbik2012,Lin2010}. The algorithm is run for $G=1~\text{to}~5$ and a two-component model is selected. Comparing the estimated component membership with the gender yields an ARI $= 0.83$. The contour plot of the fitted model in Figure~\ref{fig:contor_ais} shows that the fitted model captured the density of the data fairly well. The Gaussian mixture models and mixtures of generalized hyperbolic distributions are also applied to the AIS dataset and the summary of the performance are given in Table \ref{tab:ais}.The Gaussian mixture model selects a three component model whereas the mixtures of generalized hyperbolic distribution selects a two component model. However, the ARI of the proposed model is higher than that obtained by mixtures of generalized hyperbolic distribution. \par

                \begin{table}
            \caption{Summary of the performances of the proposed model, the Gaussian mixture model, and mixtures of generalized hyperbolic distributions on the AIS datasets.}
            \centering
            \begin{tabular*}{\textwidth}{c@{\extracolsep{\fill}} cc}
                \hline
                    &   \textbf{Estimated Groups}   &   \textbf{ARI}    \\\hline
                Proposed Algorithm & $G=2$ & 0.83\\
                GMM &   $G=3$   &   0.69    \\
                MixGHD  &   $G=2$   &   0.78    \\\hline
            \end{tabular*}
        \label{tab:ais}
        \end{table}

\begin{figure}[h!]
            \begin{minipage}{.5\textwidth}
                \centering
                \includegraphics[width=.95\linewidth,height=3 in]{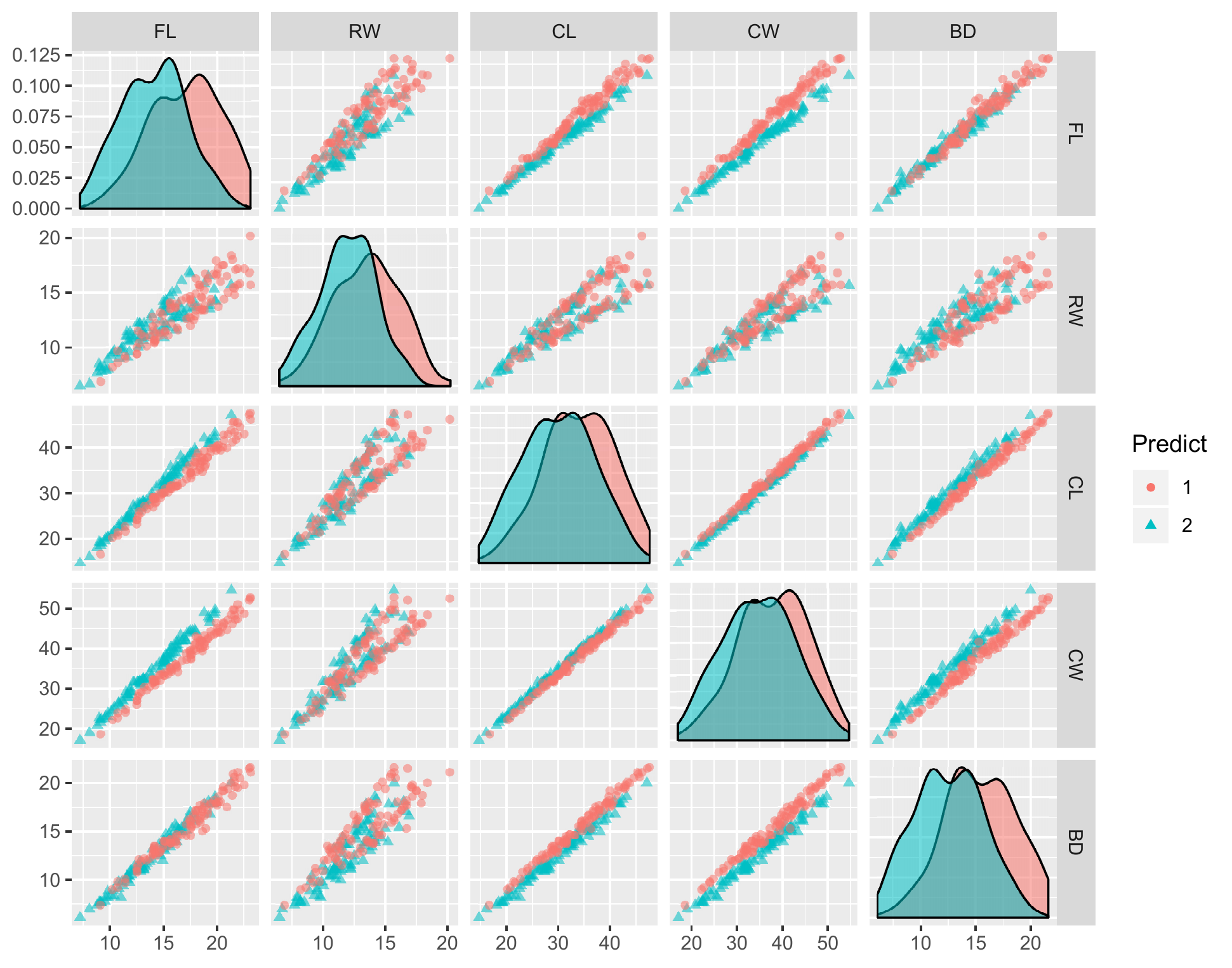}
            \end{minipage}%
            \begin{minipage}{.5\textwidth}
                \centering
                \includegraphics[width=.95\linewidth,height=3 in]{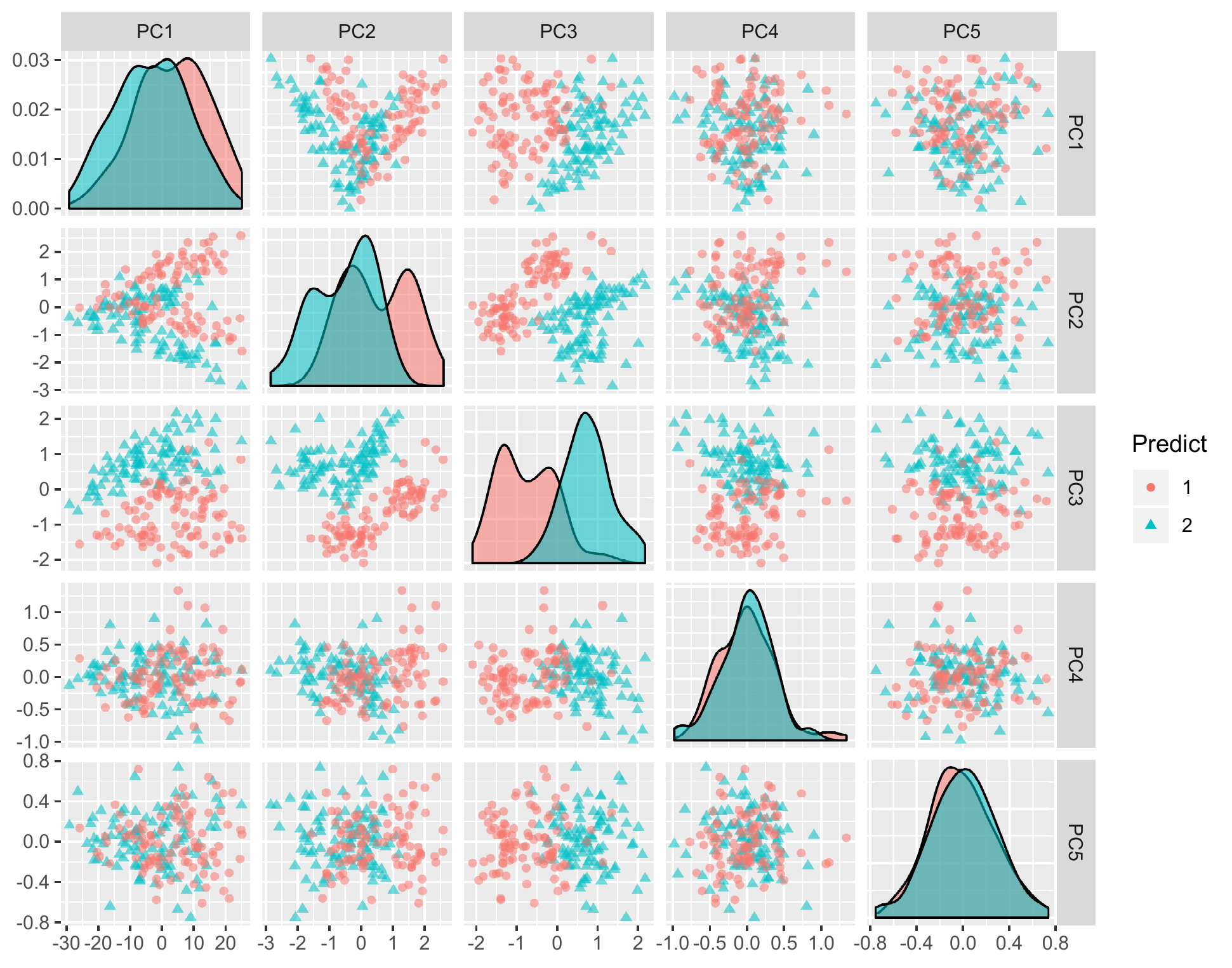}
            \end{minipage}
            \caption{Pairwise scatter plot highlighting the predicted classification of the color forms of the crabs, in terms of the original measurements (left), and in terms of the principal components (right).}
            \label{fig:crabs}
        \end{figure}

        \noindent \textbf{The Crab Dataset}:\\
        This is dataset of morphological measurements on Leptograpsus crabs, available in the \texttt{R} package \texttt{MASS} \citep{MASS}. There are 200 observations and 5 covariates in this dataset, describing 5 morphological measurements on 50 crabs each of two colour forms and both sexes. The five measurements are frontal lobe size (FL), rear width (RW), carapace length (CL), carapace width (CW), and body depth (BD) respectively. All measurements are taken in the unit of millimeters. The proposed algorithm is applied to this dataset for $G=1$ to $G=5$ and it selects a two-component model. Comparison of the estimated group membership with the two color forms of the crabs, ``B'' or ``O'' for blue or orange shows complete agreement (ARI$=1$). The pairwise scatter plots are given in Figure~\ref{fig:crabs}, where the left panel shows the original measurement variables and the right panel gives the principal components (only for visualization purposes), both colored with estimated classification of the color forms.
        The Gaussian mixture models and the mixtures of generalized hyperbolic distributions are also applied to this dataset and the performance is summarized in Table~\ref{tab:compare_crabs}.
        \begin{table}[h!]
            \centering
            \vspace*{0.1 in}

        \caption{Summary of the performances of the proposed model, the Gaussian mixture model, and mixtures of generalized hyperbolic distributions on the crabs datasets.}
            \begin{tabular*}{\textwidth}{c@{\extracolsep{\fill}} ccc}
                \hline
                &   \textbf{Model Chosen}   &   \textbf{ARI (color)}    & \textbf{ARI (gender)}\\\hline
                GMM &   $G=3$   &   0.02 & 0.72 \\
                MixGHD  &   $G=2$ & 0.00    &   0.77    \\
                Proposed Algorithm & $G=2$ & 1.00 & 0.00 \\\hline
            \end{tabular*}
            \label{tab:compare_crabs}
        \end{table}
        The mixtures of generalized hyperbolic distributions selects a two-component model (see Table \ref{tab:compare_crabs}). However, the estimated group membership are more in agreement with classifying the crabs by their sexes (ARI of 0.77).  The Gaussian mixture model on the other hand yielded a three component model where the classification obtained are also more in agreement with the classification based on the sex of the crabs (ARI of 0.72) (see Table \ref{tab:compare_crabs}).\par

\section{Conclusion}
 \label{discussion}
    In summary, we proposed a fully Bayesian approach for parameter estimation for mixtures of MNIG distributions. We also propose novel approaches to simulate from two distributions: GIG distributions and MGIG distributions. Through simulation studies, we show that the parameter estimates were very close to the true parameters and close to perfect classification was obtained. In simulation studies, the correct number of components were always selected using BIC. Using both simulated and real datasets, we show that  our approach provides competitive clustering results
compared to other clustering approaches. The contour plots using the estimated parameters show that the mixture distribution captures the shape of the dataset closely. \par

    When the true number of groups is unknown which is typically the case, the optimal number of components is chosen a posteriori using a model selection criterion. At present, the BIC is most popular but different model selection can choose different models and the BIC does not always choose the best one.  In our case, the model selection using BIC seemed to perform fairly well  in both simulated and real data analysis. However, the search for a highly effective model selection criteria remains an open problem especially when dealing with skewed data.

\section*{Appendix}
\appendix
%\section{Mathematical Details}
\section{The matrix-GIG distribution}\label{mGIGapp}
The MGIG was introduced by \cite{barndorff1982exponential}.  Let $\mathcal{V}_n$ be the Euclidean space of $n \times n$ real
symmetric matrices. Let also $\mathcal{V}_n^+$ denote the cone of
positive definite matrices in $\mathcal{V}_n$.

A random matrix  ${\bf X}$ taking its values in $\mathcal{V}_n^+$
is said to follow the $MGIG_n(-p,a,b)$ distribution if it has
density of the form
\[
f_{\bf X}({\bf x}) \propto |{\bf x}|^{-p-\frac{n+1}{2}}
\exp\left(- Tr(a{\bf x}) - Tr(b{\bf x}^{-1}) \right)
\]
The above density is well defined if $a,b \in \mathcal{V}_n^+$ and
$p \in \cal{R}$. According to \cite{Massam2006}, the density is
well defined for non positive definite matrices $a$ and $b$ if
certain conditions hold, but these cases are not of interest for
our purpose.

Similarly,  a random matrix ${\bf X}$ taking its values in
$\mathcal{V}_n^+$ is said to follow the Wishart (denoted as
$W_n(q,c)$) distribution if it has density of the form
\[
f_X(x) \propto |x|^{q-\frac{n+1}{2}} \exp\left( -Tr(cx)  \right)
\]
were $c \in \mathcal{V}_n^+$, $q > (n-1)/2$.

    \section{Mathematical Details for Posterior Distributions}\label{math_detail}

    {\small
    \begin{equation*}
    \begin{split}
   & f_{U_{ig}}(u|\Yv_i=\yv_i)
    \propto f_\Yv(\Yv|U=u)f_U(u)\\
    &\propto u^{-\frac{d+3}{2}}\exp\left\{ -\half\left(\frac{\delta_g^2}{u}+\gamma_g^2u\right) -\half (\yv_i-\muv_g - u\Deltav_g\betav_g)(u\Deltav_g)^{-1}(\yv_i-\muv_g-u\Deltav_g\betav_g)^\top\right\}\\
    &= u^{-\frac{d+3}{2}}\exp\left\{ -\half\left(\frac{\delta_g^2}{u}+\gamma_g^2u\right) -\half\left[\left((\yv_i - \muv_g)\Deltav^{-1}(\yv_i - \muv_g)^\top\right)u^{-1}+(\betav_g\Deltav_g\betav_g^\top)u\right]\right\}\\
    &= u^{-\frac{d+3}{2}}\exp\left\{ -\half\left(\frac{\delta_g^2+(\yv_i - \muv_g)\Deltav^{-1}(\yv_i - \muv_g)^\top}{u}+(\gamma_g^2+\betav_g\Deltav_g\betav_g^\top)u\right)\right\}\\
    &=u^{-\frac{d+3}{2}}\exp\left\{ -\half\left(\frac{q^2(\yv_i)}{u}+\alpha_g^2u\right)\right\}.
    \end{split}
    \end{equation*}
    Therefore, the posterior distribution of $U_{ig}|\Yv_i$ are the $\GIG$ distributions:
    \begin{equation*}
    \begin{split}
    U_{ig}|\Yv_i=\yv_i &\sim \GIG\left(\frac{d+1}{2},q^2(\yv_i),\alpha_g^2\right),\\
    U_{ig}^{-1}|\Yv_i=\yv_i &\sim \GIG\left(-\frac{d+1}{2},\alpha_g^2,q^2(\yv_i)\right).
    \end{split}
    \end{equation*}
    The contribution from $\left(\delta_g,\gamma_g\right)$ to the likelihood is given as follows:
    \begin{equation*}
    f(\delta_g,\gamma_g|\cdot) \propto \delta_g^{2\cdot\half\sum_{i}{z_{ig}}}\times\exp\left\{\delta_g\gamma_g\sum_{i}{z_{ig}}+\half\left(\gamma_g^2\sumn{z_{ig}u_{ig}} + \delta_g^2\sumn{ z_{ig}u_{ig}^{-1}}\right)\right\}.
    \end{equation*}
    Given the prior of $\delta_g^2$ of the form
    \begin{equation*}
    \delta_g^2 \sim Gamma\left(\frac{a_{g,0}^{(0)}}{2}+1, a_{g,4}^{(0)}-\frac{{a^{(0)}_{g,0}}^2}{4a_{g,3}^{(0)}}\right),
    \end{equation*}
    the posterior distribution of $\delta_g^2$ is derived as below:
    \begin{equation*}
    \begin{split}
    f(\delta_g^2|\cdot) \propto &  {\delta_g}^{2\cdot \half\sum_{i}{z_{ig}}}\cdot\exp\left\{\delta_g\gamma_g\sum_{i}{z_{ig}}+\half\delta_g^2\left(\sumn{ z_{ig}u_{ig}^{-1}}\right)\right\}\\
    &\times \left(\delta_g^2\right)^{a_{g,0}^{(0)}/2 + 1}\cdot \exp\left\{-\delta_g^{2}\left(a_{g,4}^{(0)}-\frac{{a^{(0)}_{g,0}}^2}{4a_{g,3}}\right)\right\}\\
    = & (\delta_g^2)^{a_{g,0}/2+1}\times \exp\left\{-\delta_g^2 \left(a_{g,4}^{(0)}-\frac{{a^{(0)}_{g,0}}^2}{4a_{g,3}^{(0)}}\right)\right\}.\\
    \end{split}
    \end{equation*}
    which indicates that the posterior distribution of $\delta_g^2$ is
    \begin{equation*}
    \delta_g^2 \sim Gamma\left(\frac{a_{g,0}}{2}+1, a_{g,4}-\frac{{a_{g,0}}^2}{4a_{g,3}}\right).
    \end{equation*}
    Conditional on $\delta_g$, a truncated Normal conjugate prior is assigned to $\gamma_g$. The resulting posterior distribution is given as:
    \begin{equation*}
    \gamma_g|\delta_g \sim \normal\left(\dfrac{a_{g,0}\delta_g}{2a_{g,3}},\dfrac{1}{2a_{g,3}}\right)\cdot\1v\left(\gamma_g > 0\right).
    \end{equation*}
    Conjugate joint multivariate normal prior conditional on $\Deltav_g$ was assigned to \((\muv_g,\betav_g)\), i.e.,
    \begin{equation*}
    \left.\begin{pmatrix}
    \muv_g\\
    \betav_g
    \end{pmatrix}\right\vert\Deltav_g\sim MVN
    \left[\begin{pmatrix}
    \muv_0\\
    \betav_0
    \end{pmatrix},
    \begin{pmatrix}
    \Sigmav_{\mu_0} & \Sigmav_{\mu_0,\beta_0}\\
    \Sigmav_{\mu_0,\beta_0} & \Sigmav_{\beta_0}
    \end{pmatrix}\right],
    \end{equation*}
    where \\$\muv_0 = \dfrac{2\av_{g,2}^{(0)}a_{g,3}^{(0)} - a_{g,0}^{(0)}\av_1^{(0)}}{4a_{g,3}^{(0)}a_{g,4}^{(0)} - {a^{(0)}_{g,0}}^2}$, $\betav_0 = \dfrac{\Deltav_g^{-1}(2\av_{g,1}^{(0)}a_{g,4}^{(0)} - a_{g,0}^{(0)}\av_{g,2}^{(0)})}{4a_{g,3}^{(0)}a_{g,4}^{(0)} - {a^{(0)}_{g,0}}^2}$, $\Sigmav_{\mu_0} = \dfrac{2a_{g,3}^{(0)}\Deltav_g}{4a_{g,3}^{(0)}a_{g,4}^{(0)} - {a^{(0)}_{g,0}}^2}, \\ \Sigmav_{\beta_0} = \dfrac{2a_{g,4}^{(0)}\Deltav^{-1}_g}{4a_{g,3}^{(0)}a_{g,4}^{(0)} - {a^{(0)}_{g,0}}^2}$ and $\Sigmav_{\mu_0,\beta_0} = \dfrac{-a_{g,0}^{(0)}\Id_d}{4a_{g,3}^{(0)}a_{g,4}^{(0)} - {a^{(0)}_{g,0}}^2}$.\\
    
    Therefore, the posterior should be a joint multivariate normal distribution as well of the form
    \begin{equation*}
    \left.\begin{pmatrix}
    \muv_g\\
    \betav_g
    \end{pmatrix}\right\vert\cdot\sim MVN
    \left[\begin{pmatrix}
    \muv^*\\
    \betav^*
    \end{pmatrix},
    \begin{pmatrix}
    \Sigmav_{\mu} & \Sigmav_{\mu,\beta}\\
    \Sigmav_{\mu,\beta} & \Sigmav_{\beta}
    \end{pmatrix}\right].
    \end{equation*}
    The joint density of $(\muv_g,\betav_g)$ from the likelihood is given as follows:
    \begin{equation*}
    \begin{split}
    f(\muv_g,\betav_g\mid \cdot)  \propto & \exp\left\{\left(-\betav_g^\top\muv_g\right)\sum_{i}{z_{ig}}+\betav_g^\top\sumn{z_{ig}\yv_{i}} + \muv_g^\top\Deltav_g^{-1}\sumn{z_{ig}u_{ig}^{-1}\yv_{i}}\right.\\
    & \left.\quad \quad -\half\left[\betav_g^\top\left(\sumn{u_{ig}z_{ig}}\right)\Deltav_g\betav_g\right] - \half\left[\muv_g^\top\left(\sumn{u_{ig}^{-1}z_{ig}}\right)\Deltav_g^{-1}\muv_g\right]\right\}.\\
    \end{split}
    \end{equation*}
    Hence, the marginal density of $\muv_g$ and $\betav_g$ from the posterior are given as the following:
    \begin{equation*}
    \begin{split}
    f(\muv_g|\betav_g=\betav_0,\Deltav_g) \propto & \exp\left\{-\half\left[\muv_g^\top(2a_{g,4}\Deltav_g^{-1})\muv_g - \muv_g^\top\left(\Deltav_g^{-1}\av_{g,2} - a_{g,0}\betav_g\right)\right.\right.\\
    &- \left(\Deltav_g^{-1}\av_{g,2} - a_{g,0}\betav_0\right)^\top\muv_g \\
     & \left.\left.+ \left(\Deltav_g^{-1}\av_{g,2} - a_{g,0}\betav_0\right)^\top\left(\frac{\Deltav_g}{2a_{g,4}}\right)\left(\Deltav_g^{-1}\av_{g,2} - a_{g,0}\betav_0\right)\right]\right\},\\
    f(\betav_g|\muv_g=\muv_0,\Deltav_g) \propto & \exp\left\{-\half\left[\betav_g^\top(2a_{g,3}\Deltav_g)\betav_g - \betav_g^\top\left(\av_{g,1} - a_{g,0}\muv_0\right)\right.\right.\\
     & \left.\left.\quad - \left(\av_{g,1} - a_{g,0}\muv_0\right)^\top\betav_g + \left(\av_{g,1} - a_{g,0}\muv_0\right)^\top\left(\frac{\Deltav_g^{-1}}{2a_3}\right)\left(\av_{g,1} - a_{g,0}\muv_0\right)\right]\right\}.\\
    \end{split}
    \end{equation*}
    Based on the conditional normal property of the multivariate normal variables,
    \begin{equation*}
    \begin{split}
    \left(\muv_g|\betav_g = \betav_0, \Deltav_g, \cdot \right)& \sim MVN\left(\muv^* + \Sigmav_{\mu,\beta}\Sigmav_{\beta}^{-1}(\betav_0 - \betav^*),\Sigmav_{\mu}-\Sigmav_{\mu,\beta}\Sigmav_{\beta}^{-1}\Sigmav_{\mu,\beta}\right)\quad \text{and}\\
    \left(\betav_g|\muv_g=\muv_0,\Deltav_g,\cdot\right)&\sim MVN\left(\betav^*+\Sigmav_{\mu,\beta}\Sigmav_{\mu}^{-1}(\muv_0-\muv^*), \Sigmav_{\beta}-\Sigmav_{\mu,\beta}\Sigmav_{\mu}^{-1}\Sigmav_{\mu,\beta}\right),
    \end{split}
    \end{equation*}
    we can derive that
    \begin{equation*}
    \muv^* = \dfrac{2\av_{g,2}a_{g,3}- a_{g,0}\av_1}{4a_{g,3}a_{g,4} - {a_{g,0}}^2} \quad \quad \betav^* = \dfrac{\Deltav_g^{-1}(2\av_{g,1}a_{g,4} - a_{g,0}\av_{g,2})}{4a_{g,3}a_{g,4} - {a_{g,0}}^2},
    \end{equation*}
    \begin{equation*}
    \Sigmav_{\mu} = \dfrac{2a_{g,3}\Deltav_g}{4a_{g,3}a_{g,4} - {a_{g,0}}^2},\quad \Sigmav_{\beta} = \dfrac{2a_{g,4}\Deltav^{-1}_g}{4a_{g,3}a_{g,4} - {a_{g,0}}^2},\quad \Sigmav_{\mu,\beta} = \dfrac{-a_{g,0}\Id_d}{4a_{g,3}a_{g,4}-{a_{g,0}}^2}.
    \end{equation*}
     An Inverse-Wishart prior is given to $\Deltav_g$, the resulting posterior is a Matrix Generalized Inverse-Gaussian (MGIG) distribution. The derivation is given as below:
    \begin{equation*}
    \begin{split}
    &f(\Deltav_g|\cdot) \propto  |\Deltav_g|^{-\frac{\sum_i{z_{ig}}}{2}}\times \exp Tr\left\{-\betav_g\left(\half \sumn{u_{ig}z_{ig}}\right)\betav_g^\top \Deltav_g - \sumn{\frac{(\yv_i - \muv_g)(\yv_{i} - \muv_g)^\top}{2u_i}}\Deltav_g^{-1}\right\}\\
    & \times |\Deltav_g|^{-\frac{\nu_0+d+1}{2}}\times \exp Tr\left\{-\half \Lambda_0\Deltav_g^{-1}\right\}\\
    = & |\Deltav_g|^{-\frac{t_{g,0}+\nu_0+d+1}{2}}\times \exp Tr\left\{-\half \betav_g\betav_g^\top (2t_{g,3})\Deltav_g - \half\left(\sumn{z_{ig}u_{ig}^{-1}(\yv_i - \muv_g)(\yv_{i} - \muv_g)^\top}+\Lambda_0\right)\Deltav_g^{-1}\right\}\\
    =& |\Deltav_g|^{-\frac{t_{g,0}+\nu_0+d+1}{2}}\times \exp Tr\left\{-\half \betav_g\betav_g^\top (2t_{g,3})\Deltav_g - \half\left({S_0}_g+\Lambda_0\right)\Deltav_g^{-1}\right\}.\\
    \end{split}
    \end{equation*}
    Therefore, the posterior of $\Deltav_g$ is
    \begin{equation*}
    \Deltav_g\sim MGIG_d\left(-\frac{\nu_0+t_{0g}}{2}, \betav_g 2t_{3g}\betav_g^\top, {S_0}_g+\Lambda_0\right).
    \end{equation*}
    }
% \bibliographystyle{JASA}
%\bibliography{projbib}
%\newpage

\section{Credible Intervals for Simulation Study 1}\label{supp_fig}

    \begin{figure*}[!h]
       \vspace{-1.5in}    
        \centering
        \includegraphics[width=.8\linewidth]{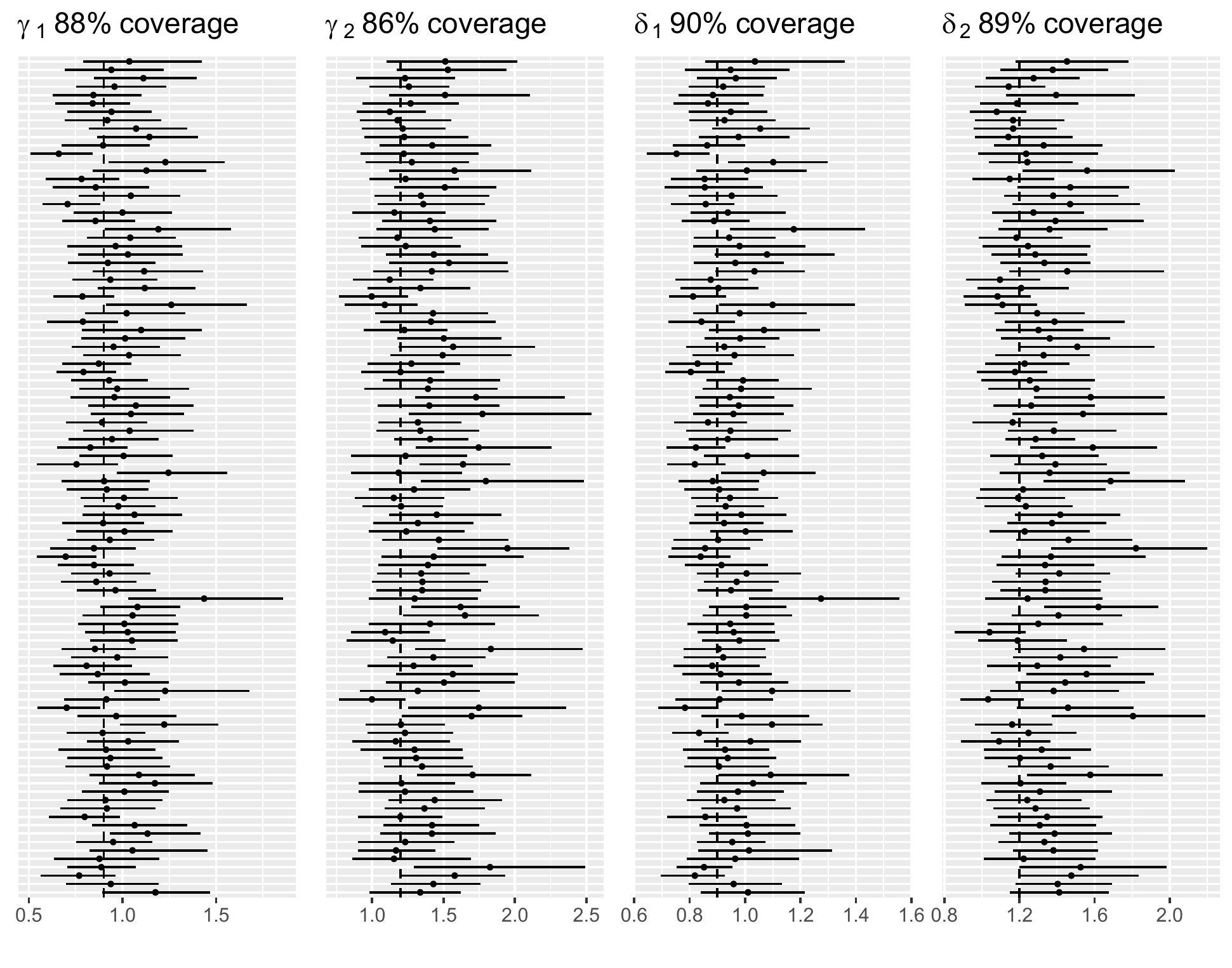}
        \vspace{-2.5in}    
        
        \includegraphics[width=.8\linewidth]{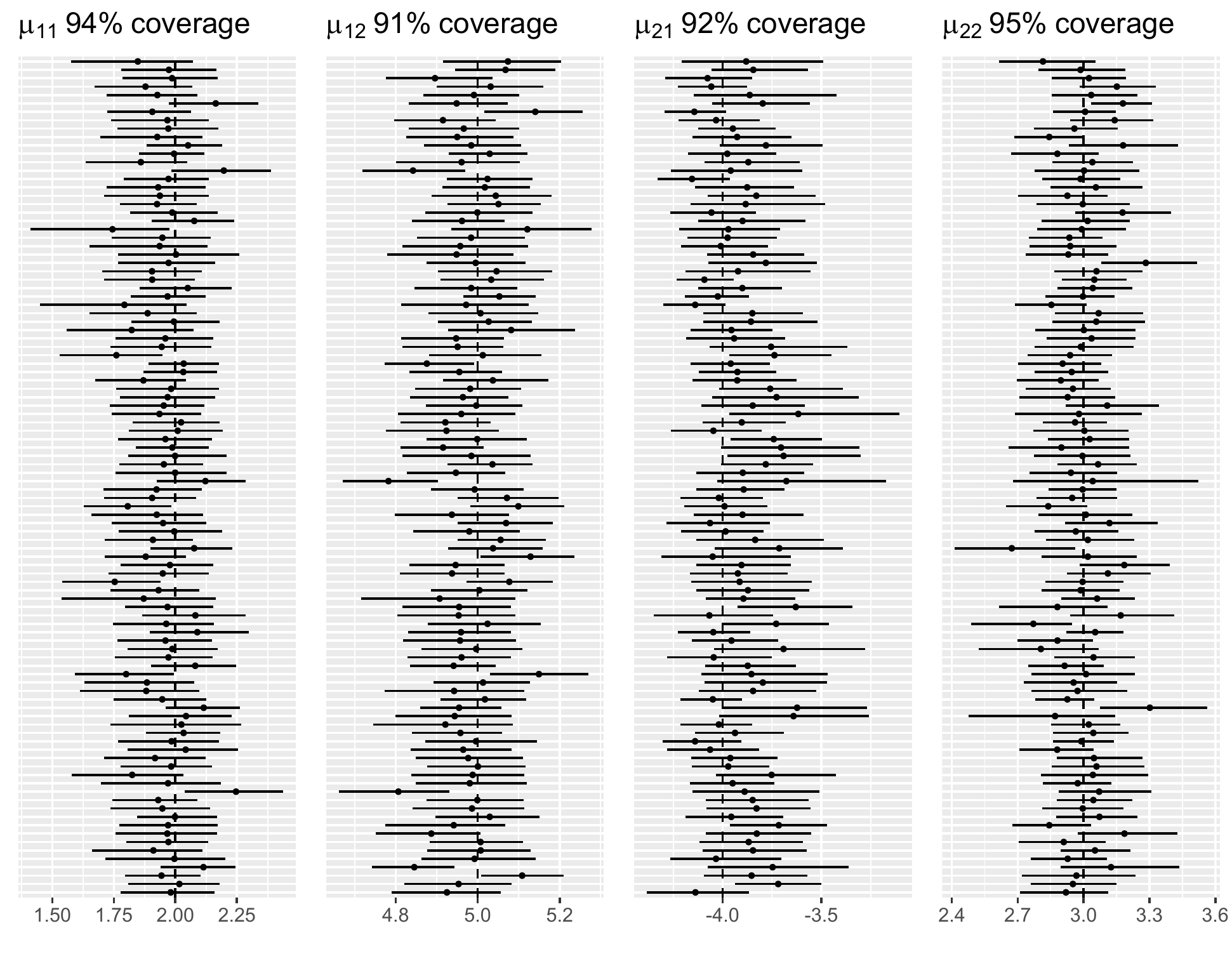}
    \end{figure*}
    \begin{figure*}[h]
      \vspace{-1.5in}    
        \centering
        \includegraphics[width=.8\linewidth]{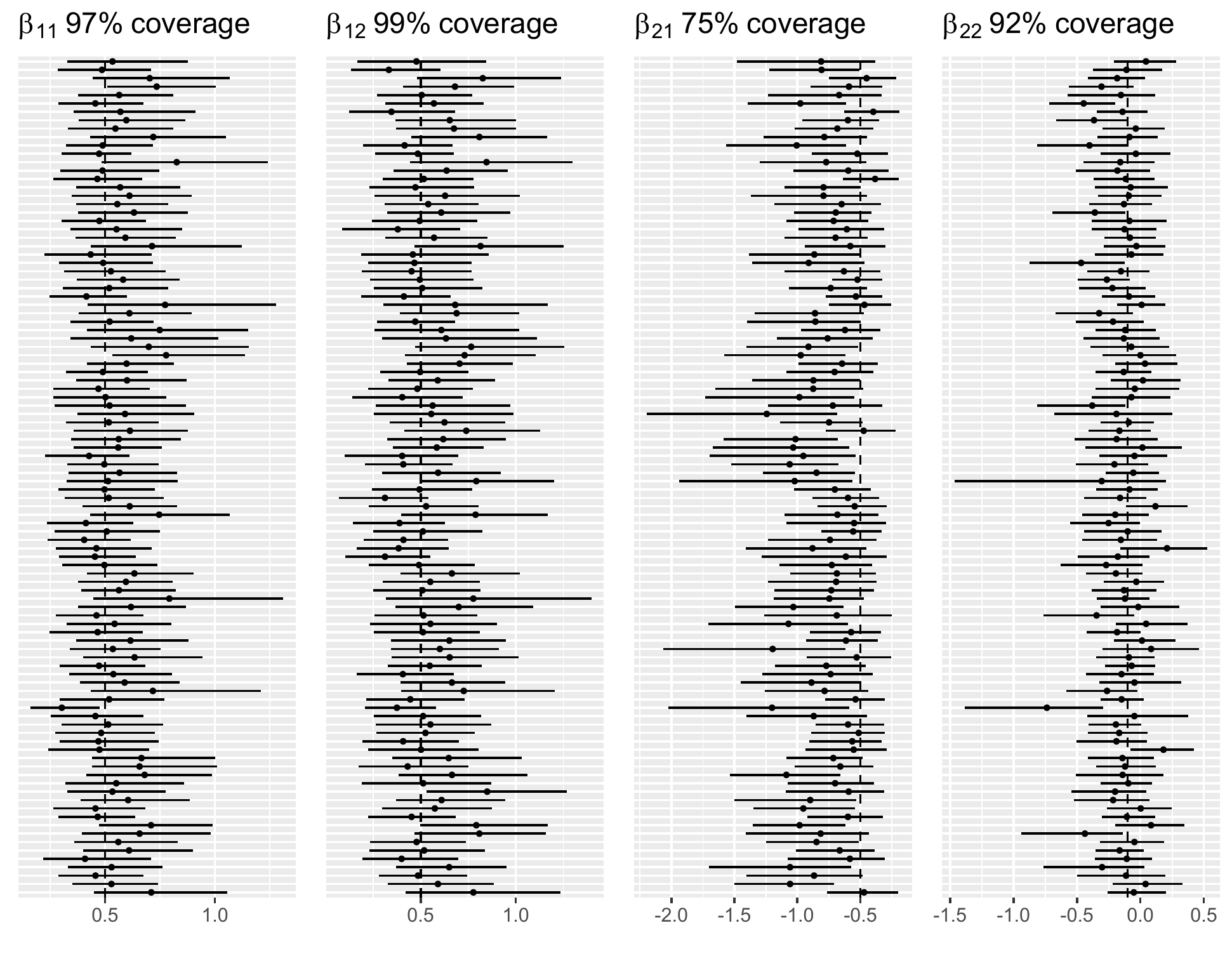}
          \vspace{-2.5in}  
          
        \includegraphics[width=.8\linewidth]{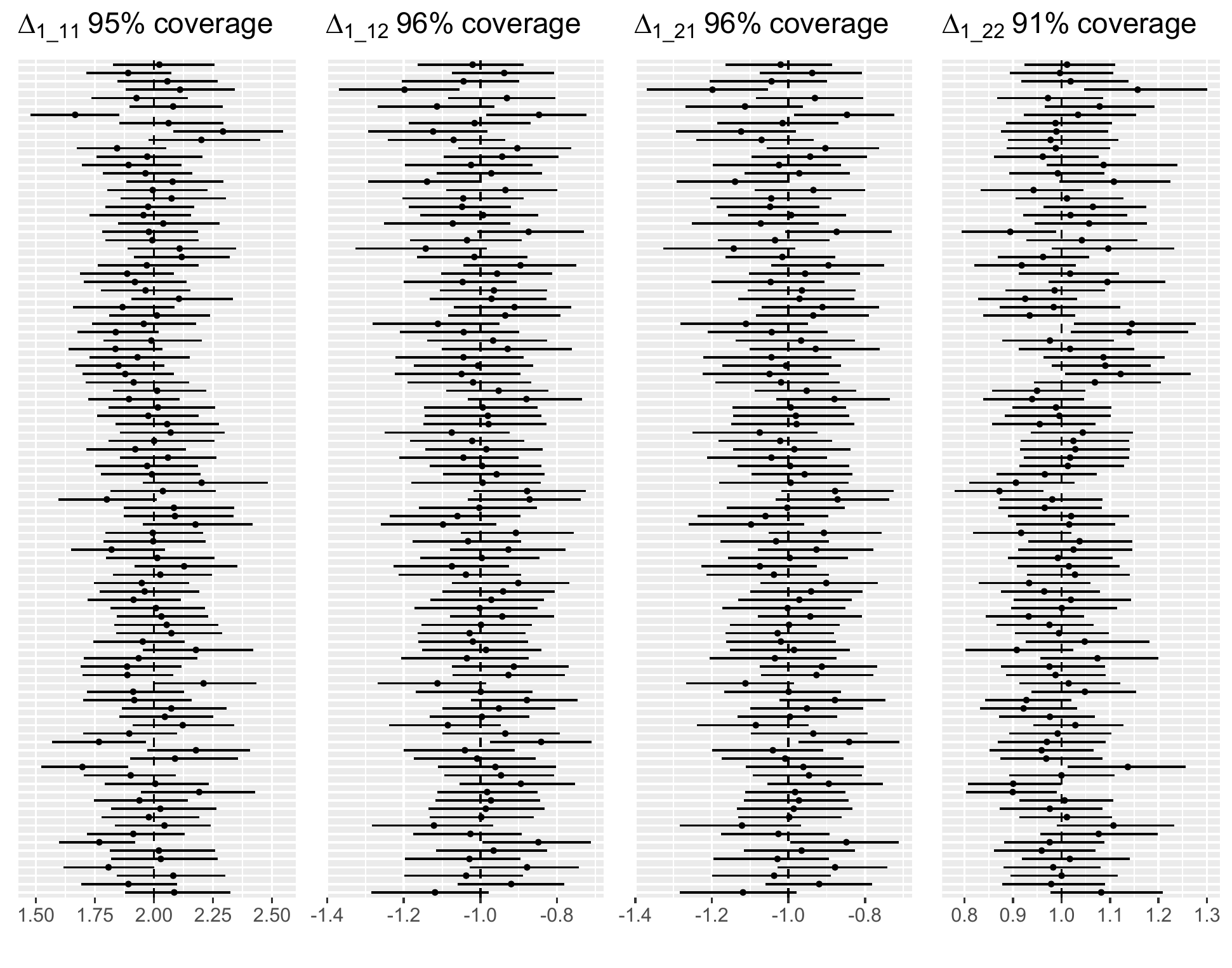}
    \end{figure*}
    \begin{figure*}[h]
        \centering
        \includegraphics[width=.9\linewidth]{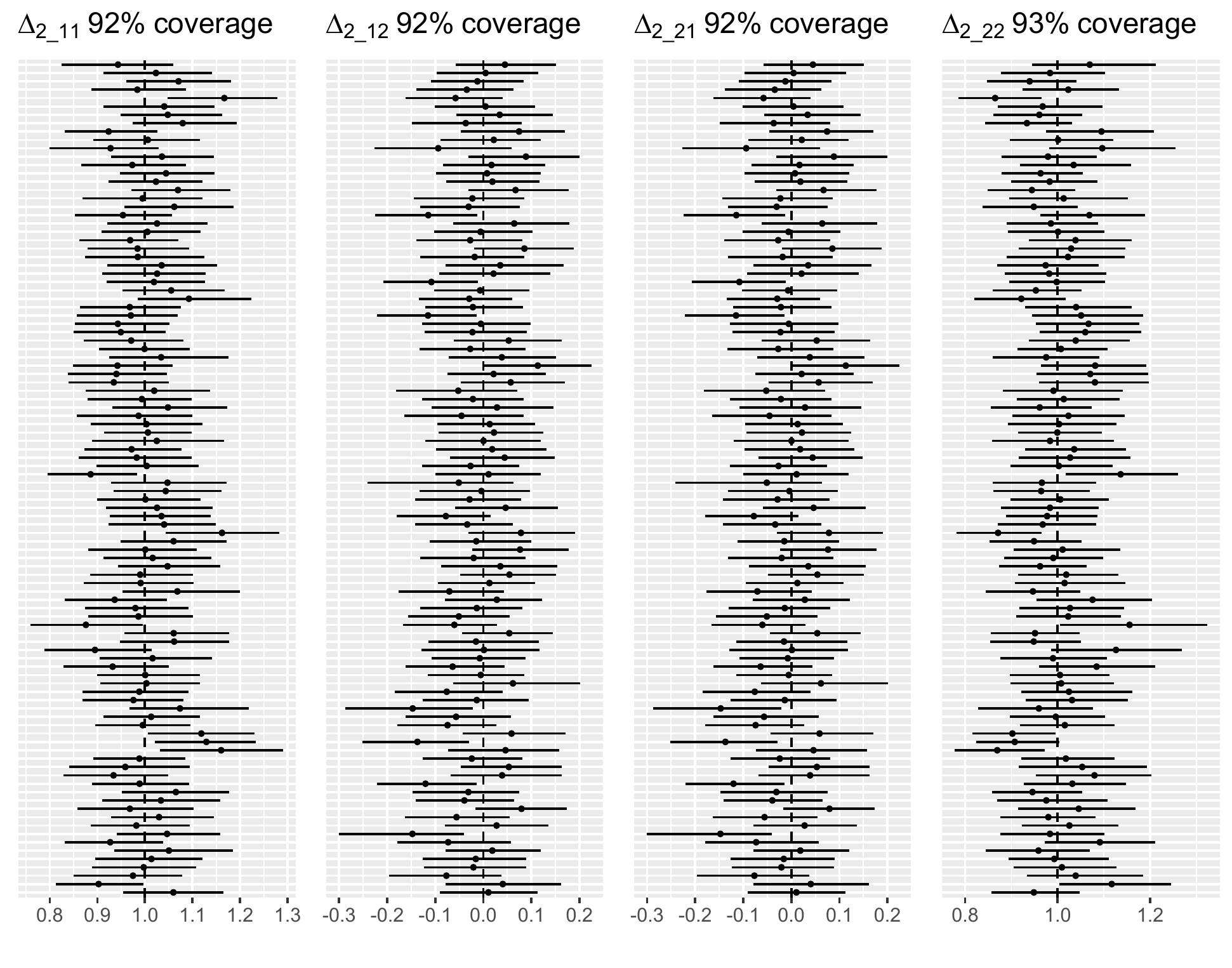}
        \caption{95\% Credible Intervals for estimated parameters for all 100 runs for Simulation Study 1.}
        \label{fig:parameter_ci}
    \end{figure*}

\end{document}